\RequirePackage{ifpdf}
\ifpdf
    \documentclass[pdftex,envcountsect,envcountsame,orivec]{llncs}
\else
    \documentclass[dvips,envcountsect,envcountsame,orivec]{llncs}
\fi

\usepackage{subfig}
\usepackage{xspace}
\usepackage{textgreek}
\usepackage{latexsym,amssymb,amsmath,amsfonts}
\usepackage{tabularx}
\usepackage{paralist}
\usepackage{todonotes}
\usepackage{fancyvrb}
\usepackage{listings}
\usepackage{booktabs}
\usepackage{array,multirow,graphicx}

\usepackage{enumerate}
\usepackage[latin1]{inputenc}
\usepackage[pdftitle   = {{BPM}},
            pdfauthor  = {{Calvanese -Ghilardi - Gianola - Montali - Rivkin}},
            pdfsubject = {{}}]{hyperref}
\ifpdf \usepackage{url} \else \usepackage{breakurl} \fi
\usepackage{comment}
\usepackage{graphicx}
\usepackage{tikz}
\usetikzlibrary{matrix,arrows, calc,fit,shapes, trees, positioning, decorations.markings, patterns,calc,backgrounds} 
\usepackage{bpmn-events}
\usepackage{bpmn-gateways}
\usepackage{wrapfig}

\usepackage{amsmath,amssymb}

\usepackage[linesnumberedhidden,ruled,vlined]{algorithm2e}
\usepackage{graphicx}

\usepackage{mathptmx}
\DeclareMathAlphabet{\mathcal}{OMS}{cmsy}{m}{n}

\title{Formal Modeling and SMT-Based Parameterized Verification of Data-Aware BPMN}
\subtitle{(Extended Version)}

\author{%
Diego Calvanese$^1$ , Silvio Ghilardi$^2$,  Alessandro Gianola$^1$, \\ Marco Montali$^1$, Andrey Rivkin$^1$ 
}

\institute{%
$^1$Faculty of Computer Science, Free University of Bozen-Bolzano (Italy)\\
$^2$Dipartimento di Matematica, Universit\`a degli Studi di Milano (Italy) 
}
 
\newcommand{\D}{\ensuremath{\mathcal{D}}}

\newcommand{\M}{\ensuremath{\mathcal{M}}}

\renewcommand{\P}{\ensuremath{\mathcal{P}}}

\newcommand{\R}{\ensuremath{\mathcal{R}}}

\newcommand{\V}{\ensuremath{\mathcal{V}}}

\newcommand{\set}[1]{\{#1\}}                      

\newcommand{\tup}[1]{\langle #1\rangle}            

\newcommand{\inlinetitle}[1]{\smallskip\noindent\textbf{#1.}\xspace}

\newcolumntype{C}{>{\centering\arraybackslash}X}

\makeatletter
\g@addto@macro\normalsize{%
\setlength{\abovecaptionskip}{-2pt}
\setlength{\belowcaptionskip}{-10pt}
\setlength\abovedisplayskip{3pt}
\setlength\belowdisplayskip{3pt}
\setlength\abovedisplayshortskip{3pt}
\setlength\belowdisplayshortskip{3pt}
}
\makeatother

\newcounter{dummy} 
\newcounter{dummy1} 
\newcounter{dummy2}
\newcounter{dummy3} 
\newcounter{dummy4}
\newcounter{dummy5} 
\newcounter{dummy6}

\usepackage[thmmarks,amsmath]{ntheorem}
\theorempreskip{1pt}
\theorempostskip{1pt}

\theoremseparator{.}
\theorembodyfont{\itshape}
\theoremsymbol{$\triangleleft$}
\newtheorem{theorem}[dummy]{Theorem}
\newtheorem{lemma}[dummy1]{Lemma}
\newtheorem{definition}[dummy2]{Definition}
\newtheorem{proposition}[dummy3]{Proposition}

\theorembodyfont{\normalfont}
\newtheorem{example}[dummy4]{Example}
\newtheorem{remark}[dummy5]{Remark}

\theoremstyle{nonumberplain}
\theoremheaderfont{\itshape}
\theorembodyfont{\normalfont}

\theoremseparator{.}
\theoremsymbol{\ensuremath{\dashv}}
\newtheorem{proof}[dummy6]{Proof}

\qedsymbol{\ensuremath{\dashv}}

\newcommand{\dab}{DAB\xspace}
\newcommand{\dabs}{DABs\xspace}

\newcommand{\true}{\constant{true}\xspace}
\newcommand{\false}{\constant{false}\xspace}

\newcommand{\safe}{\texttt{SAFE}\xspace}

\newcommand{\mcmt}{\textsc{mcmt}\xspace}

\renewcommand{\int}{\ensuremath {\mathcal I}}

\newcommand{\relname}[1]{\ensuremath{\mathit{#1}}\xspace}

\newcommand{\constant}[1]{\texttt{#1}}
\newcommand{\sorts}[1]{#1}
\newcommand{\sort}[1]{\sorts{\mathsf{#1}}}

\newcommand{\nullv}{\texttt{undef}}

\definecolor{deepblue}{HTML}{0C3B80}
\definecolor{deepgreen}{HTML}{2EA601}
\definecolor{lightOrange}{HTML}{FFA03C}
\definecolor{darkOrange}{HTML}{F1800A}
\definecolor{lightBlue}{HTML}{0174CD}
\definecolor{greenF}{HTML}{2CBB5C}
\definecolor{cyan}{HTML}{86A6D5}

\tikzstyle{sortnode} = [
  rectangle,
  rounded corners=5pt,
  minimum width=18mm,
  minimum height=6mm,
  very thick,
]

\tikzstyle{functnode} = [
  rectangle,
  minimum width=1cm,
  minimum height=5mm,
]

\tikzstyle{idnode} = [
  sortnode,
  draw=deepblue,
  fill=cyan!50,
  text=deepblue,  
]

\tikzstyle{valnode} = [
  sortnode,
  densely dashed,
  draw=violet,
  fill=magenta!50,
  text=violet,
]

\tikzstyle{f} = [
  thick,
]

\tikzstyle{fd} = [
  f,
  -angle 90,
]

\tikzstyle{relation}=[rectangle split, rectangle split parts=#1, rectangle split part align=base, draw, anchor=center, align=center, text height=3mm, font=\bfseries, text centered]

\newcommand{\SET}{\xspace\mathtt{SET}\xspace}
\newcommand{\INSERT}{\mathtt{INSERT}\xspace}
\newcommand{\INTO}{\mathtt{INTO}\xspace}
\newcommand{\FROM}{\mathtt{FROM}\xspace}

\newcommand{\MOVE}{\mathtt{DEL}\xspace}

\newcommand{\SETTING}{\mathtt{AND~SET}\xspace}
\newcommand{\UPDATE}{\mathtt{UPDATE}\xspace}
\newcommand{\IF}{\xspace\mathtt{IF}\xspace}
\newcommand{\THEN}{\xspace\mathtt{THEN}\xspace}
\newcommand{\ELSE}{\xspace\mathtt{ELSE}\xspace}
\newcommand{\nvar}[1]{\mathit{?}#1\xspace}

\tikzstyle{sortnode} = [
  rectangle,
  rounded corners=5pt,
  minimum width=18mm,
  minimum height=6mm,
  very thick,
]

\tikzstyle{functnode} = [
  rectangle,
  minimum width=1cm,
  minimum height=5mm,
]

\tikzstyle{idnode} = [
  sortnode,
  draw=deepblue,
  fill=cyan!50,
  text=deepblue,  
]

\tikzstyle{artnode} = [
  sortnode,
  draw=deepblue,
  fill=yellow!50,
  text=deepblue,  
]

\tikzstyle{valnode} = [
  sortnode,
  densely dashed,
  draw=violet,
  fill=magenta!50,
  text=violet,
]

\tikzstyle{f} = [
  thick,
]

\tikzstyle{fd} = [
  f,
  -angle 90,
]

\tikzstyle{relation}=[rectangle split, rectangle split parts=#1, rectangle split part align=base, draw, anchor=center, align=center, text height=3mm, font=\bfseries, text centered]

\definecolor{deepblue}{HTML}{0C3B80}
\definecolor{deepgreen}{HTML}{2EA601}
\definecolor{lightOrange}{HTML}{FFA03C}
\definecolor{darkOrange}{HTML}{F1800A}
\definecolor{lightBlue}{HTML}{0174CD}
\definecolor{greenF}{HTML}{2CBB5C}
\definecolor{cyan}{HTML}{86A6D5}

\tikzstyle{task} = [
  minimum height=1cm,
  minimum width = 1.4cm,
  rectangle,
  fill=white,
  thick,
  draw,
  rounded corners,
  align=center,
  every task, 
  font=\footnotesize
]

\tikzstyle{sequence} = [
  ->,
  >=triangle 45,
  every sequence,
  thick
]

\tikzstyle{guard} = [
  rectangle,
  draw=none,
  fill=white,
  inner sep=.5mm,
  minimum height=.8mm,
  minimum width=.8mm
]

\tikzstyle{lbl} = [text width=4cm]
\newcommand{\tname}[1]{\textsf{#1}} 

\makeatletter
\pgfarrowsdeclare{center*}{center*}
{
  \pgfarrowsleftextend{+-.5\pgflinewidth}
  \pgfutil@tempdima=0.4pt%
  \advance\pgfutil@tempdima by.2\pgflinewidth%
  \pgfarrowsrightextend{4.5\pgfutil@tempdima}
}
{
  \pgfutil@tempdima=0.4pt%
  \advance\pgfutil@tempdima by.2\pgflinewidth%
  \pgfsetdash{}{+0pt}
  \pgfpathcircle{\pgfqpoint{4.5\pgfutil@tempdima}{0bp}}{4.5\pgfutil@tempdima}
  \pgfusepathqfillstroke
}

\pgfarrowsdeclare{centero}{centero}
{
  \pgfarrowsleftextend{+-.5\pgflinewidth}
  \pgfutil@tempdima=0.4pt%
  \advance\pgfutil@tempdima by.2\pgflinewidth%
  \pgfarrowsrightextend{4.5\pgfutil@tempdima}
}
{
  \pgfutil@tempdima=0.4pt%
  \advance\pgfutil@tempdima by.2\pgflinewidth%
  \pgfsetdash{}{+0pt}
  \pgfpathcircle{\pgfqpoint{4.5\pgfutil@tempdima}{0bp}}{4.5\pgfutil@tempdima}
  \pgfusepathqstroke
}
\makeatother

\begin{document}


\maketitle


\begin{abstract}
We propose DAB -- a data-aware extension of BPMN where the process operates over case and persistent data (partitioned into a read-only database called catalog and a read-write database called repository). The model trades off between expressiveness and the possibility of supporting parameterized verification of safety properties on top of it. 
Specifically, taking inspiration from the literature on verification of artifact systems, we study verification problems where safety properties are checked irrespectively of the content of the read-only catalog, and accepting the potential presence of unboundedly many tuples in the catalog and repository. 
We tackle such problems using an array-based backward reachability procedure fully implemented in MCMT -- a state-of-the-art array-based SMT model checker. Notably, we prove that the procedure is sound and complete for checking safety of \dab{s}, and single out additional conditions that guarantee its termination and, in turn, show decidability of checking safety.
\end{abstract}

\section{Introduction}

In recent years, increasing attention has been given to multi-perspective models of business processes that strive to capture the interplay between the process and data dimensions~\cite{Reic12}. Conventional finite-state verification techniques only work in this setting if data are abstractly represented, e.g., as finite sate machines \cite{MuRH07} or process annotations \cite{SiST11}. If data are instead tackled in their full generality, verifying whether a process meets desired temporal properties (e.g., is safe) becomes highly undecidable, and cannot be directly attacked using conventional finite-state model checking techniques \cite{CaDM13}. This triggered a flourishing research on the formalization and the boundaries of verifiability of data-aware processes, focusing mainly on data- and artifact-centric models~\cite{CaDM13,DeHL18}.  Recent results in this stream of research \cite{DeLV16,CGGMR19} come with two strong advantages. First, they consider the relevant setting where the running process evolves a set of relations (henceforth called a data \emph{repository}) containing data objects that may have been injected from the external environment (e.g., due to user interaction), or borrowed from a read-only relational database with constraints (henceforth called \emph{catalog}). The repository acts as a working memory and a log for the process. Notably, it may accumulate unboundedly many tuples resulting from complex constructs in the process, such as while loops whose repeated activities insert new tuples in the repository (e.g., the applications sent by candidates in response to a job offer).  
The catalog stores background, contextual facts that do not change during the process execution, such as the catalog of product types, the usernames and passwords of registered customers in an order-to-cash process. 
In this setting, verification is studied parametrically to the catalog, so as to ensure that the process works as desired irrespectively of the specific read-only data stored therein. 
This is crucial to verify the process under robust conditions, also considering that actual data may not yet be available at modeling time.  
The second main advantage of these techniques is that they tame the infinity of the state space to be verified with a symbolic approach, in turn paving the way for the development of feasible implementations \cite{verifas,DeHV14}, or for the exploitation of state-of-the-art symbolic model checkers for infinite-state systems \cite{CGGMR19,mcmt}.

In a parallel research line more conventional, activity-centric approaches, such as the de-facto standard BPMN, have been extended towards data support, mainly focusing on conceptual modeling and enactment \cite{MPFW13,DOET17,COWZ18}, but not on verification. 
At the same time, several formalisms have been brought forward to capture multi-perspective processes based on Petri nets enriched with various forms of data: from data items locally carried by tokens \cite{RVFE11,Las16}, to 
case data with different data types \cite{DeFM18}, and/or persistent relational data manipulated with the full power of FOL/SQL \cite{DDG16,MonR17}. 
While these formalisms qualify well to directly capture data-aware extensions of BPMN (e.g., \cite{MPFW13,DOET17}), they suffer of two main limitations. On the foundational side, they require to specify the data present in the read-only storage, and only allow boundedly many tuples (with an a-priori known bound) to be stored in the read-write ones. On the applied side, they do not lend themselves to be symbolically verified and have not yet led to the development of actual verifiers. 

This leads us to the main question tackled by this paper: \emph{how to extend BPMN towards data support, guaranteeing the applicability of the existing parameterized verification techniques and the corresponding actual verifiers, so far studied only in the artifact-centric setting?} 
We answer this question by considering the framework of \cite{CGGMR19} and the verification of safety properties (i.e., properties that must hold in every state of the analyzed system). 
Specifically, our \emph{first contribution} is a data-aware extension of BPMN called \dab, which supports case data, as well as persistent relational data partitioned into a read-only catalog and a read-write repository. Case and persistent data are used to express conditions in the process as well as task preconditions; tasks, in turn, change the values of the case variables and insert/update/delete tuples into/from the repository.

The resulting framework is similar, in spirit, to the BAUML approach \cite{EsST15}, which relies on UML and OCL instead of BPMN as we do here. While \cite{EsST15} approaches verification via a translation to first-order logic with time, we follow a different route, by encoding \dabs into the array-based artifact system framework from \cite{CGGMR19}. 
Thanks to this encoding, we can effectively verify safety properties of \dabs using the well-established \mcmt (\emph{Model Checker Modulo Theories}) model checker~\cite{ijcar08,lmcs}. \mcmt implements a symbolic backward reachability procedure that 
relies on state-of-the-art Satisfiability Modulo Theories (SMT) solvers, and that has been widely employed to verify infinite-state \emph{array-based systems}.

Using the encoding above, we provide our \emph{second contribution}: we show that this backward reachability procedure  is sound and complete when it comes to checking safety of \dab{s}. 
In this context, soundness means that whenever the procedure terminates the returned answer is correct, whereas completeness means that if the process is unsafe then the procedure will always discover it.

The fact that the procedure is sound and complete does not guarantee that it will always terminate. This brings us to the \emph{third and last contribution} of this paper: 
we introduce further conditions that, by carefully controlling the interplay between the process and data components, guarantee the termination of the procedure. Such conditions are expressed as syntactic restrictions over the \dab under study, thus providing a concrete, BPMN-grounded counterpart of the conditions imposed in \cite{verifas,CGGMR19}.
By exploiting the encoding from \dabs to array-based artifact systems, and the soundness and completeness of backward reachability, we derive that 
checking safety for the class of \dabs satisfying these conditions is decidable.

To show that our approach goes end-to-end from theory to actual verification, we finally report some preliminary experiments demonstrating how \mcmt checks safety of \dabs. 

This paper is the extended version of~\cite{BPM19}. 
Full proofs of our technical results and the files of the experiments with \mcmt can be found in \cite{CGGMR19-techrep-dab-multicase}.

\newcommand{\types}{\mathcal{S}}
\newcommand{\atype}{S}

\newcommand{\cat}{\mathit{Cat}}
\newcommand{\caseid}{\mathit{CType}}
\newcommand{\repo}{\mathit{Repo}}
\newcommand{\cvars}{\mathit{X}}

\newcommand{\vtypes}{\types_{v}}
\newcommand{\idtypes}{\types_{id}}
\newcommand{\datadom}{\mathbb{D}}
\newcommand{\vdatadom}{\datadom_v}
\newcommand{\iddatadom}{\datadom_{id}}

\newcommand{\cvar}[1]{\mathbf{#1}}
\renewcommand{\nvar}[1]{\mathit{#1}}

\newcommand{\dotcomp}[1]{\mathsf{#1}}
\newcommand{\dotexpr}[2]{#1.\dotcomp{#2}}

\newcommand{\reln}[1]{\dotexpr{\relname{#1}}{name}}
\newcommand{\relid}[1]{\dotexpr{\relname{#1}}{id}}
\newcommand{\relattrs}[1]{\dotexpr{\relname{#1}}{attrs}}

\newcommand{\dcat}[1]{\dotexpr{#1}{cat}}
\newcommand{\drepo}[1]{\dotexpr{#1}{repo}}
\newcommand{\dcvars}[1]{\dotexpr{#1}{cvars}}
\newcommand{\dctype}[1]{\dotexpr{#1}{ctype}}

\newcommand{\freevars}[1]{\mathit{free}(#1)}
\newcommand{\getdatavars}[1]{\mathit{normvars}(#1)}
\newcommand{\getcasevars}[1]{\mathit{casevars}(#1)}

\newcommand{\tpre}{G}
\newcommand{\tpost}{E}

\newcommand{\getpre}[1]{\dotexpr{#1}{pre}}
\newcommand{\getpost}[1]{\dotexpr{#1}{eff}}

\newcommand{\updatespec}[1]{\texttt{#1}}

\newcommand{\dhr}{\D^h}

\newcommand{\bldist}{-1mm}
\newcommand{\outfdist}{4mm}
\tikzstyle{outflow} = [sequence,->,densely dotted]
\tikzstyle{smalltask} = [rectangle,draw,rounded corners=5pt,minimum height=2.5em,minimum width=3em]
\tikzstyle{block} = [task,densely dotted]
\tikzstyle{smallblock} = [smalltask,densely dotted]
\tikzstyle{legend} = [font=\footnotesize]

\tikzstyle{state} = [
  rectangle,
  draw,
  rounded corners=10pt,
  minimum width=15mm,
  minimum height=7mm]

\newcommand{\sidle}{\constant{idle}}
\newcommand{\senabled}{\constant{enabled}}
\newcommand{\sactive}{\constant{active}}
\newcommand{\scompleted}{\constant{compl}}
\newcommand{\scomplerror}{\constant{error}}

\section{Data-aware BPMN}



We start by describing our formal model of data-aware BPMN processes (\dab{s}). We focus here on private, single-pool processes, analyzed considering a single case, similarly to soudness analysis in workflow nets \cite{Aalst97}.\footnote{The interplay among multiple cases is also crucial. The technical report \cite{CGGMR19-techrep-dab-multicase} already contains an extension of the framework presented here, in which multiple cases are modeled and verified.} Incoming messages are therefore handled as pure nondeterministic events. The model combines a wide range of (block-structured) BPMN control-flow constructs with task, event-reaction, and condition logic that inspect and modify persistent as well as case data. 
Given the aim of our approach, recall that if something is not supported in the language, it is because it would hamper soundness and completeness of SMT-based (parameterized) verification.

First, some preliminary notation. 
 We consider a set $\types = \vtypes \uplus \idtypes$ of (semantic) \emph{types}, consisting of \emph{primitive types} $\vtypes$ accounting for data objects, and \emph{id types} $\idtypes$ accounting for identifiers. We assume that each type $\atype \in \types$ comes with a (possibly infinite) domain $\datadom_\atype$, a special constant $\nullv_\atype \in \datadom_\atype$ to denote an undefined value in that domain, and a type-wise equality operator $=_\atype$. We omit the type and simply write $\nullv$ and $=$ when clear from the context. We do not consider here additional type-specific predicates (such as comparison and arithmetic operators for numerical primitive types); these will be added in future work. In the following, we simply use \emph{typed} as a shortcut for $\types$-\emph{typed}. We also  denote by $\datadom$ the overall domain of objects and identifiers (i.e., the union of all domains in $\types$). 
We consider a countably infinite set $\V$ of typed variables. Given a variable or object $x$, we may explicitly indicate that $x$ has type $\atype$ by writing $x:\atype$. We omit types whenever clear  or irrelevant. We compactly indicate a possibly empty tuple $\tup{x_1,\ldots,x_n}$ of variables as $\vec{x}$, and with slight abuse of notation, we write $\vec{x} \subseteq \vec{y}$ if all variables in $\vec{x}$ also appear in $\vec{y}$.

\subsection{The Data Schema}
\label{sec:data-schema}
Consistently with the BPMN standard, we consider two main forms of data: \emph{case data}\footnote{These are called \emph{data objects} in BPMN, but we prefer to use the term \emph{case data} to avoid name clashes with the formal notions.}, instantiated and manipulated on a per-case basis; \emph{persistent data} (cf.~data store references in BPMN), accounting for global data that are accessed by all cases. For simplicity, case data are defined at the whole process level, and are directly visible by all tasks and subprocesses (without requiring the specification of input-output bindings and the like).

To account for persistent data, we consider relational databases. We describe relation schemas by using the \emph{named perspective}, i.e., by assigning a dedicated typed attribute to each component (i.e., column) of a relation schema. Also for an attribute, we use the notation $a:\atype$ to explicitly indicate its type. 
\begin{definition} 
\label{def:relation-schema}
A \emph{relation schema} is a pair $R=\tup{N,A}$, where:
\begin{inparaenum}[\it (i)]
\item $N = \reln{R}$ is the relation \emph{name};
\item $A = \relattrs{R}$ is a nonempty tuple of attributes.  
\end{inparaenum}
\end{definition}
We call $|A|$ the \emph{arity} of $R$. We assume that distinct relation schemas use distinct names, blurring the distinction between the two notions (i.e., we set $\reln{R} = R$). We also use the predicate notation $R(A)$ to represent a relation schema $\tup{R,A}$. An example of a relation schema is given by $\relname{User}(Uid{:}\sort{Int},Name{:}\sort{String})$, where the first component represents the id-number of a user, whereas the second component is the string formed by her name.

\inlinetitle{Data schema}
First of all, we define the \emph{catalog}, i.e., a read-only, persistent storage of data that is not modified during the execution of the process. 
Such a storage could contain, for example, the catalog of product types and the set of registered customers and their addresses in an order-to-cash scenario.
\begin{definition}
\label{def:catalog}
A \emph{catalog} $\cat$ is a set of relation schemas satisfying the following requirements:
\begin{compactdesc}
  \item[(single-column primary key)] Every relation schema $R$ is such that the first attribute in $\relattrs{R}$ has type in $\idtypes$, and denotes the \emph{primary key} of the relation; we refer to such attribute using the dot notation $\relid{R}$.
  \item[(non-ambiguity of primary keys)] for every pair $R_1$ and $R_2$ of \emph{distinct} relation schemas in $\cat$, we have that the types of $\relid{R_1}$ and $\relid{R_2}$ are different.
  \item[(foreign keys)] for every relation schema $R \in \cat$ and non-id attribute $a \in \relattrs{R} \setminus \relid{R}$ with type $\atype \in \idtypes$, there exists a relation schema $R_2 \in \R$ such that the type of $\relid{R_2}$ is $\atype$; $a$ is hence a \emph{foreign key} referring to $R_2$.
\end{compactdesc}  
\end{definition}

\begin{example}
  \label{ex:relation-schema}
  Consider a simplified example of a job hiring process in a company. 
  To represent information related to the process we make use of the  $\cat$ consisting of the following relation schemas: 
\begin{compactitem}[$\bullet$]
\item $\relname{JobCategory}(Jcid{:}\sort{jobcatID})$ contains the different job categories available in the company (e.g., programmer, analyst, and the like) - we just store here the identifiers of such categories;
\item $\relname{User}(Uid{:}\sort{userID},Name{:}\sort{StringName},Age{:}\sort{NumAge})$ stores data about users registered to the company website, and who are potentially interested in job positions offered by the company.
\end{compactitem}
Each case of the process is about a job. Jobs are identified by the type $\sort{jobcatID}$.
\end{example}

We now define the data schema of a BPMN process, which combines a catalog with:
\begin{inparaenum}[\it (i)]
\item a persistent data \emph{repository}, consisting of updatable relation schemas possibly referring to the catalog;
\item a set of \emph{case variables}, constituting local data carried by each process case.  
\end{inparaenum}



\begin{definition}
\label{def:data-component}
A \emph{data schema} $\D$ is a tuple $\tup{\cat,\repo,\cvars}$, where
\begin{inparaenum}[\it (i)]
\item $\cat = \dcat{\D}$ is a \emph{catalog},
\item $\repo = \drepo{\D}$ is a set of relation schemas called \emph{repository}, and
\item $\cvars = \dcvars{\D} \subset \V$ is a finite set of typed variables called \emph{case variables},
\end{inparaenum} 
such that:
\begin{compactitem}[$\bullet$]
  \item for every relation schema $R \in \repo$ and every attribute $a \in \relattrs{R}$ whose type is $S \in \idtypes$, there exists $R \in \cat $ such that the type of $\relid{R}$ is $S$;
\item for every case variable $\cvar{x} \in \cvars$ whose type is $S \in \idtypes$, there exists $R \in \cat$ such that the type of $\relid{R}$ is $S$.
\end{compactitem}
\end{definition}
We use bold-face to distinguish a case variable $\cvar{x}$ from a ``normal" variable $x$. 
It is worth noting that relation schemas in the repository are not equipped with an explicit primary key, and thus they cannot reference each other, but may contain foreign keys pointing to the catalog or the case identifiers. \emph{This is essential towards soundness and completeness of SMT-based verification of \dab{s}}. It will be clear how tuples can be inserted and removed from the repository once we will introduce updates.

\begin{example}
  \label{ex:data-schema}
To manage key information about the applications submitted for the job hiring, the company employs a repository that consists of one relation schema: 
\[
\relname{Application}(
\begin{array}[t]{@{}l@{}}
Jcid{:}\sort{JobcatID},Uid{:}\sort{UserID},Eligible{:}\sort{Bool})
\end{array}
\]  $\sort{NumScore}$ is a finite-domain type containing $100$ scores in the range
      $[\constant{1},\constant{100}]$. For readability, we use the usual comparison predicates for variables of type $\sort{NumScore}$: this is syntactic sugar and does not require to introduce datatype predicates in our framework. 
%
Since each posted job is created using a dedicated portal, its corresponding data do not have to be stored persistently and thus can be maintained just for a given case. At the same time, some specific values have to be moved from a specific case to the repository and vice-versa. This is done by resorting to the following case variables $\dcvars{\D}$:
\begin{inparaenum}[\it (i)]
\item $\cvar{jcid}:\sort{jobcatID}$ references a job type from the catalog, matching the type of job associated to the case;
\item $\cvar{uid}:\sort{userID}$ 
references the identifier 
 of a user who is applying for the job associated to the case;
\item $\cvar{result}:\sort{Bool}$ indicates whether the user identified by $\cvar{uid}$ is eligible for winning the position or not;
\item $\cvar{qualif}:\sort{Bool}$ indicates whether the user identified by $\cvar{uid}$ qualifies for directly getting the job (without the need of carrying out a comparative evaluation of all applicants); 
\item $\cvar{winner}:\sort{userID}$ contains the identifier of the applicant winning the position.
\end{inparaenum}
\end{example}

At runtime, a \emph{data snapshot} of a data schema consists of three components:
\begin{compactitem}[$\bullet$]
\item An immutable \emph{catalog instance}, i.e., a fixed set of tuples for each relation schema contained therein, so that the primary and foreign keys are satisfied. 
\item An assignment mapping case variables to corresponding data objects. 
\item A \emph{repository instance}, i.e., a set of tuples forming a relation for each schema contained therein, so that the foreign key constraints pointing to the catalog are satisfied. Each tuple is associated to a distinct primary key that is not explicitly accessible.
\end{compactitem}

\inlinetitle{Querying the data schema}
To inspect the data contained in a snapshot, we need suitable query languages operating over the data schema of that snapshot. 
We start by considering boolean \emph{conditions} over (case) variables. These conditions will be attached to choice points in the process.

\begin{definition}
  \label{def:condition}
  A \emph{condition} is a formula of the form $\varphi ::= (x=y) \mid \neg \varphi \mid \varphi_1 \land \varphi_2$, where $x$ and $y$ are variables from $\V$ or constant objects from $\datadom$. 
\end{definition}
We make use of the standard abbreviation $\varphi_1 \lor \varphi_2 = \neg (\neg \varphi_1 \land \neg \varphi_2)$. 

We now extend conditions to also access the data stored in the catalog and repository, and to ask for data objects subject to constraints. We consider the well-known language of unions of conjunctive queries with atomic negation, which correspond to unions of select-project-join SQL queries with table filters. 
%

\begin{definition}
A \emph{conjunctive query with filters} over a data component $\D$ is a formula of the form
  $Q ::= \varphi \mid R(x_1,\ldots,x_n) \mid \neg R(x_1,\ldots,x_n) \mid Q_1 \land Q_2$, where $\varphi$ is a condition with only \emph{atomic} negation, $R\in  \dcat{\D} \cup \drepo{\D}$ is a relation schema of arity $n$, and $x_1,\ldots,x_n$ are variables from $\V$ (including $\dcvars{\D}$) or constant objects from $\datadom$. We denote by $\freevars{Q}$ the set of variables occurring in $Q$ that are \emph{not} case variables in $\dcvars{\D}$.  
\end{definition}
For example, a conjunctive query $\relname{JobCategory}(jt)\land jt\neq \constant{HR}$ lists all the job categories available in the company, apart from HR.
\begin{definition}
\label{def:guard}
  A \emph{guard} $G$ over a data component $\D$ is an expression of the form $q(\vec{x}) \leftarrow \bigvee_{i=1}^n Q_i$, where:
  \begin{inparaenum}[\it (i)]
 \item $q(\vec{x})$ is the \emph{head} of the guard with \emph{answer variables} $\vec{x}$;
  \item each $Q_i$ 
  is a conjunctive query with filters over $\D$; 
  \item for some $i \in \set{1,\ldots,n}$, $\vec{x} \subseteq \freevars{Q_i}$.
  \end{inparaenum} 
  We denote by $\getcasevars{G} \subseteq \dcvars{\D}$ the set of case variables used in $G$, and by $\getdatavars{G} = \bigcup_{i \in \set{1,\ldots,n}} \freevars{Q_i}$ the other variables used in $G$. 
\end{definition}
To distinguish guard heads from relations, we write the former in camel case, while the latter shall always begin with capital letters.  
\begin{definition}
  A \emph{guard} $G$ over a data component $\D$ is \emph{repo-free} if none of its atoms queries a relation schema from $\drepo{\D}$.
\end{definition}

Notice that \emph{going beyond this guard query language} (e.g., by introducing universal quantification) \emph{would hamper the soundness and completeness of SMT-based verification over the resulting \dab{s}}. We will come back to this important aspect in the conclusion. 

As anticipated before, this language can be seen as a standard query language to retrieve data from a snapshot, but also as a mechanism to constrain the combinations of data objects that can be injected into the process. E.g., a simple guard $\relname{input}(y{:}\sort{string},z{:}\sort{string}) \rightarrow y \neq z$ returns all pairs of strings that are different from each other. Picking an answer in this (infinite) set of pairs can be interpreted as a (constrained) user input where the user decides the values for $y$ and $z$. 


%

\subsection{Tasks, Events, and Impact on Data}
\label{sec:update-logic}
We now formalize how the process can access and update the data component when executing a task or reacting to the trigger of an external event. 

\inlinetitle{The update logic}
We start by discussing how data maintained in a snapshot can be subject to change while executing the process.
\begin{definition}
\label{def:update-spec}
Given a data schema $\D$, an \emph{update specification} $\alpha$ is a pair $\tup{\tpre,\tpost}$, where:
\begin{inparaenum}[\it (i)]
\item $\tpre = \getpre{\alpha}$ is a guard over $\D$ of the form $q(\vec{x}) \leftarrow Q$, called \emph{precondition}; 
\item $\tpost= \getpost{\alpha}$ is an \emph{effect rule} that changes the snapshot of $\D$, as described next.
\end{inparaenum} 
Each effect rule has one of the following forms:
\begin{compactdesc}
\item[(Insert\&Set)] $\INSERT~\vec{u}~\INTO~R~\SETTING~\cvar{x}_1 = v_1, \ldots, \cvar{x}_n = v_n$, where: 
\begin{inparaenum}[\it (i)] 
\item  $\vec{u}, \vec{v}$ are variables in $\vec{x}$ or constant objects from $\datadom$; 
\item $\vec{\cvar{x}} \in \dcvars{\D}$ are distinct case variables; 
\item $R$ is a relation schema from $\drepo{\D}$ whose arity (and types) match $\vec{u}$. 
\end{inparaenum}
Either the $\INSERT$ or $\SET$ parts may be omitted, obtaining a pure  \textnormal{\textbf{Insert rule}} or  \textnormal{\textbf{Set rule}}.
\item[(Delete\&Set)] $\MOVE~\vec{u}~\FROM~R~\SETTING~\cvar{x}_1 = v_1, \ldots, \cvar{x}_n = v_n$, where:
 \begin{inparaenum}[\it (i)] 
\item  $\vec{u},\vec{v}$ are variables in $\vec{x}$ or constant objects from $\datadom$; 
\item  $\vec{\cvar{x}} \in \dcvars{\D}$; 
\item $R$ is a relation schema from $\drepo{\D}$ whose arity (and types) match $\vec{u}$. 
\end{inparaenum}
As in the previous rule type, the $\SETTING$ part may be omitted, obtaining a pure (repository) \textnormal{\textbf{Delete rule}}.
\item[(Conditional update)] $\UPDATE~R(\vec{v})~\IF~\psi(\vec{u},\vec{v})~\THEN~\eta_1~\ELSE~\eta_2$, where:
 \begin{inparaenum}[\it (i)] 
\item $\vec{u}$ is a tuple containing variables in $\vec{x}$ or constant objects from $\datadom$; 
\item $\psi$ is a repo-free guard (called \emph{filter});
\item $R$ is a relation schema from $\drepo{\D}$;
\item $\vec{v}$ is a tuple of new variables, i.e., such that $\vec{v} \cap (\vec{u} \cup \dcvars{\D}) = \emptyset$;
\item $\eta_i$ is either an atomic formula of the form $R(\vec{u}')$
 with $\vec{u}'$ a tuple of elements from $\vec{x} \cup \datadom \cup \vec{v}$, or a nested $\IF\ldots\THEN\ldots\ELSE$.
\end{inparaenum}
\end{compactdesc}
\end{definition}

We now comment on the semantics of update specifications. An update specification $\alpha$ is executable in a given data snapshot if there is at least one answer to the precondition $\getpre{\alpha}$ in that snapshot. If this is the case, then the process executor(s) can nondeterministically decide which answer to pick so as to \emph{bind} the answer variables of $\getpre{\alpha}$ to corresponding data objects in $\datadom$. This confirms the interpretation discussed in Section~\ref{sec:data-schema} for which  the answer variables of $\getpre{\alpha}$ can be seen as \emph{constrained user inputs} in case multiple bindings are available. 

Once a specific binding for the answer variables is selected, the corresponding effect rule $\getpost{\alpha}$, instantiated using that binding, is issued. How this affects the current data snapshot depends on which effect rule is adopted.

If  $\getpost{\alpha}$ is an insert\&set rule, the binding is used to \emph{simultaneously}
insert a tuple in one of the repository relations, and update some of the case variables -- with the implicit assumption that those not explicitly mentioned in the $\SET$ part  maintain their current values. 
Since repository relations do not have an explicit primary key, two possible semantics can be attached to the insertion of a tuple $\vec{u}$ in the instance of a repository relation $R$:
\begin{compactdesc}
\item[(multiset insertion)] Upon insertion, $\vec{u}$ gets an  implicit, fresh primary key. The insertion then always results in the genuine addition of the tuple to the current instance of $R$, even in the case where the tuple already exists there.
\item[(set insertion)] In this case, $R$ comes not only with its implicit primary key, but also with an additional, genuine key constraint defined over a subset $K \subseteq \relattrs{R}$ of its attributes. Upon insertion, if there already exists a tuple in the current instance of $R$ that agrees with $\vec{u}$ on $K$, then that tuple is \emph{updated} according to $\vec{u}$. If no such tuple exists, then as in the previous case $\vec{u}$ gets implicitly assigned to a fresh primary key, and inserted into the current instance of $R$. By default, if no explicit key is defined over $R$, then the entire set of attributes $\relattrs{R}$ is considered as a key, consequently enforcing a \emph{set semantics} for insertion.
\end{compactdesc}

\begin{example}
\label{ex:updates-1}
  We continue the job hiring example, by considering two update specifications of type insert\&set. When a new case is created, the first update is about indicating what is the category of job associated to the case. This is done through the update specification $\updatespec{InsJobCat}$, where $\getpre{\updatespec{InsJobCat}}$ selects a job category from the corresponding catalog relation, while $\getpost{\updatespec{InsJobCat}}$ assigns the selected job category to the case variable $\cvar{jcid}$:
$$
  \begin{array}{@{}rl@{}}
  \getpre{\updatespec{InsJobCat}} \triangleq &
      \relname{getJobType}(c) \leftarrow \relname{JobCategory}(c)\\
  \getpost{\updatespec{InsJobCat}} \triangleq &
      \SET~\cvar{jcid} = c
  \end{array}
$$
  When the case receives an application, the user id is picked from the corresponding $\mathit{User}$ via the update specification $\updatespec{InsUser}$, where:
  $$
    \begin{array}{@{}rl@{}}
      \getpre{\updatespec{InsUser}} \triangleq & 
        \relname{getUser}(u) 
        \leftarrow 
        \relname{User}(u, n, a)\\
      \getpost{\updatespec{InsUser}} \triangleq &
         \SET~\cvar{uid} = u\\ 
    \end{array}
  $$
  A different usage of precondition, resembling a pure external choice, is the update specification $\updatespec{CheckQual}$ to handle a quick evaluation of the candidate and check whether she has such a high profile qualifying her to directly get an offer:
  $$
    \begin{array}{@{}rl@{}}
      \getpre{\updatespec{CheckQual}} \triangleq & 
        \relname{isQualified}(q:\sort{Bool}) 
        \leftarrow 
        \true \\
      \getpost{\updatespec{CheckQual}} \triangleq &
         \SET~\cvar{qualif} = q\\
    \end{array}
  $$
As an example of insertion rule, we consider the situation where the candidate whose id is currently stored in the case variable $\cvar{uid}$ has not been directly judged as qualified. She is consequently subject to a more fine-grained evaluation of her application, resulting in a score that is then registered in the repository (together with the applicant data). This is done via the $\updatespec{EvalApp}$ specification:
  $$
    \begin{array}{@{}rl@{}}
      \getpre{\updatespec{EvalApp}} \triangleq & 
        \relname{getScore}(s:\sort{NumScore}) 
        \leftarrow 
        1 \leq s \land s \leq 100 \\
      \getpost{\updatespec{EvalApp}} \triangleq &
       \INSERT~\tup{\cvar{jcid},\cvar{uid},s,\nullv}~\INTO~\relname{Application}\\
    \end{array}
  $$
Here, the insertion indicates an $\nullv$ eligibility, since it will be assessed in a consequent step of the process. 
   
  Notice that, by adopting the \emph{multiset insertion semantics}, the same user may apply multiple times for the same job (resulting multiple times as applicant). With a \emph{set insertion semantics}, we could enforce the uniqueness of the application by declaring the second component (i.e., the user id) of $\relname{Application}$ as a key.
  \end{example}

If $\getpost{\alpha}$ is a delete\&set rule, then the executability of the update is subject to the fact that the tuple $\vec{u}$ selected by the binding and to be removed from $R$, is actually present in the current instance of $R$. If so, the binding is used to \emph{simultaneously} delete $\vec{u}$ from $R$ and update some of the case variables -- with the implicit assumption that those not explicitly mentioned in the $\SET$ part maintain their current values. 

Finally, a conditional update rule applies, tuple by tuple, a bulk operation over the content of $R$. For each tuple in $R$, if it passes the filter associated to the rule, then the tuple is updated according to the $\THEN$ part, whereas if the filter evaluates to false, the tuple is updated according to the $\ELSE$ part.

\begin{example}
\label{ex:updates-2}
  Continuing with our running example, we now consider the update specification $\updatespec{MarkE}$ handling the situation where no candidate has been directly considered as qualified, and so the eligibility of all received (and evaluated) applications has to be assessed. Here we consider that each application is eligible if and only if its evaluation resulted in a score greater than $80$. Technically, $\getpre{\updatespec{MarkE}}$ is a true precondition, and:
  $$
    \getpost{\updatespec{MarkE}} \triangleq \begin{array}[t]{@{}l@{}}
       \UPDATE~\relname{Application}(jc,u,s,e)\\
       \IF s > 80~\THEN~\relname{Application}(jc,u,s,\true)\\
       \ELSE~\relname{Application}(jc,u,s,\false)\\
      \end{array} 
  $$  
  If there is at least one eligible candidate, she can be selected as a winner using the $\updatespec{SelWinner}$ update specification, which deletes the selected winner tuple from $\relname{Application}$, and transfers its content to the corresponding case variables (also ensuring that the $\cvar{winner}$ case variable is set to the applicant id). Technically:
  $$
    \begin{array}{@{}rl@{}}
      \getpre{\updatespec{SelWinner}} \triangleq & 
        \relname{getWinner}(jc,u,s,e) 
        \leftarrow 
          \begin{array}[t]{@{}l@{}}
            \relname{Application}(jc,u,s,e)\\
            \land e = \true\\
          \end{array}\\
      \getpost{\updatespec{SelWinner}} \triangleq &
       \MOVE~\tup{jc,u,s,e}~\FROM~\relname{Application}\\
       &\SETTING~\cvar{jcid}= jc, \cvar{uid} = u,\cvar{winner}=u, \cvar{result}=e, \cvar{qualif}=\false
    \end{array}
  $$
  Deleting the tuple is useful in the situation where the selected winner may refuse the job, and consequently should not be considered again if a new winner selection is carried out. To keep such tuple in the repository, one would just need to remove the $\MOVE$ part from $\getpost{\updatespec{SelWinner}}$.
  \end{example}

\inlinetitle{The task/event logic}
We now substantiate how the update logic is used to specify the task/event logic within a \dab process. The first important observation, not related to our specific approach, but inherently present whenever the process control flow is enriched with relational data, is that update effects manipulating the repository must be executed in an atomic, non-interruptible way. This is essential to ensure that insertions/deletions into/from the repository are applied on the same data snapshot where the precondition is checked. Breaking simultaneity would lead to nondeterministic interleave with other update specifications potentially operating over the same portion of the repository.  
This is why in our approach we consider two types of task: \emph{atomic} and \emph{nonatomic}. 

Each atomic task/catching event is associated to a corresponding update specification. In the case of tasks, the specification precondition indicates under which circumstances the task can be enacted, and the specification effect how enacting the task impacts on the underlying data snapshot. In the case of events, the specification precondition constrains the data payload that comes with the event (possibly depending on the data snapshot, which is global and therefore accessible also from the perspective of an external event trigger), and the specification effect how reacting to a triggered event impacts on the underlying data snapshot. More concretely, this is realized according to the following lifecycle.

The task/event is initially $\sidle$, i.e., quiescent. When the progression of a case reaches an $\sidle$ task/event, such a task/event becomes $\senabled$. An $\senabled$ task/event may nondeterministically fire depending on the choice of the process executor(s). Upon firing, a binding satisfying the precondition of the update specification associated to the task/event is selected, consequently grounding and applying the corresponding effect. At the same time, the lifecycle moves from $\senabled$ to $\scompleted$. Finally, a $\scompleted$ task/event triggers the progression of its case depending on the process-control flow, simultaneously bringing the task/event back to the $\sidle$ state (which would then make it possible for the task to be executed again later, if the process control-flow dictates so). 

The lifecycle of a nonatomic task diverges in two crucial respects. First of all, upon firing it moves from $\senabled$ to $\sactive$, and later on nondeterministically from $\sactive$ to $\scompleted$ (thus having a duration). The precondition of its update specification is checked and bound to one of the available answers when the task becomes $\sactive$, while the corresponding effect is applied when the task becomes $\scompleted$. Since these two transitions occur asynchronously, to avoid the aforementioned transactional issues we assume that the effect operates, in this context, only on case variables (and not on the repository).

 \newcommand{\legendw}{8cm}
\newcommand{\bdist}{5mm}

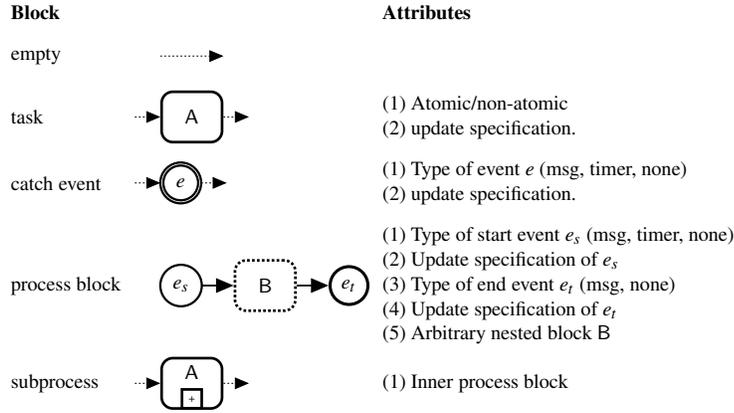
\begin{figure}[t]
\centering
\resizebox{\textwidth}{!}{
\begin{tikzpicture}[x=1.8cm,y=.9cm, thick]

\matrix[  nodes={node distance=\bldist},
            rectangle,
            nodes in empty cells,
            row sep=.5mm,column sep=1mm,
            very thick,
            column 1/.style={anchor=west},
            column 2/.style={anchor=west},
            column 3/.style={anchor=west},
            >=latex,->,
            ampersand replacement=\&
          ] (declarematrix) {
\node{\textbf{Block}};
\&
\&
\node{\textbf{Attributes}};
\\
\node{empty};
\&
  \node[smalltask,white] (task) {};
  \draw[outflow,thin] (task.west) -- (task.east);
\&
\node{};
\\
\node{task};
\&
  \node[smalltask] (task) {\textsf{A}};
  \draw[outflow,thin] ($(task.west)-(\outfdist,0)$) -- (task);
  \draw[outflow,thin] (task) -- ($(task.east)+(\outfdist,0)$);
\&
\node{
\begin{tabular}{@{}p{\legendw}@{}}
(1) Atomic/non-atomic\\
(2) update specification.
\end{tabular}
};
\\
\node{catch event};
\&
  \node[IntermediateEvent,minimum size=.6cm] (task) {$e$};
  \draw[outflow,thin] ($(task.west)-(\outfdist,0)$) -- (task);
  \draw[outflow,thin] (task) -- ($(task.east)+(\outfdist,0)$);
\&
\node{
\begin{tabular}{@{}p{\legendw}@{}}
(1) Type of event $e$ (msg, timer, none)\\
(2) update specification.
\end{tabular}
};
\\
\node{process block};
\&
\node[StartEvent,minimum size=.6cm] (start) {$e_s$};
\node[smallblock,right=\bdist of start] (task) {\textsf{B}};
\draw[sequence,->] (start) -- (task);
\node[EndEvent,minimum size=.6cm,right= 5mm of task] (end) {$e_t$};
\draw[sequence,->]  (task) -- (end);
\&
\node{
\begin{tabular}{@{}p{\legendw}@{}}
(1) Type of start event $e_s$ (msg, timer, none)\\
(2) Update specification of $e_s$\\
(3) Type of end event $e_t$ (msg, none)\\
(4) Update specification of $e_t$\\
(5) Arbitrary nested block $\textsf{B}$
\end{tabular}
};
\\
\node{subprocess};
\&
\node[smalltask] (task) at (0,0) {};
\node[below=0mm of task.north,anchor=north] {\textsf{A}};
\node[
  rectangle,
  anchor=south,
  above=0mm of task.south,
  draw,
  minimum size=.5mm,
] {\tiny +};
  \draw[outflow,thin] ($(task.west)-(\outfdist,0)$) -- (task);
  \draw[outflow,thin] (task) -- ($(task.east)+(\outfdist,0)$);
\&
\node{
\begin{tabular}{@{}p{\legendw}@{}}
(1) Inner process block
\end{tabular}
};
\\
};
\end{tikzpicture}
}
\caption{\dab Basic blocks}
\label{fig:basic-blocks}
\end{figure}

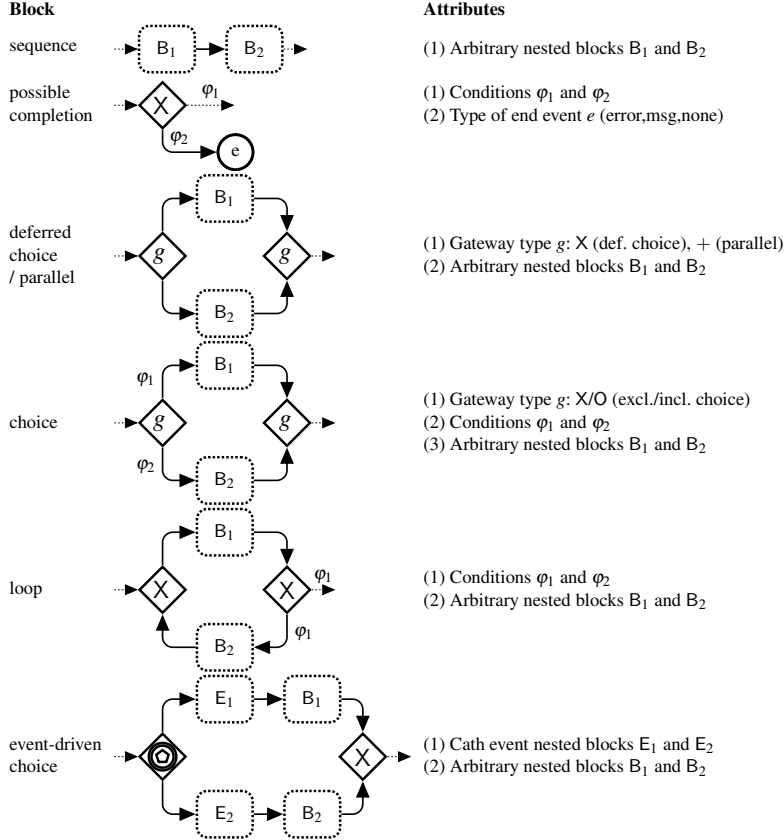
\begin{figure}[t]
\centering
\resizebox{\textwidth}{!}{
\begin{tikzpicture}[x=1.8cm,y=.9cm, thick]

\matrix[  nodes={node distance=\bldist},
            rectangle,
            nodes in empty cells,
            row sep=.5mm,column sep=1mm,
            very thick,
            column 1/.style={anchor=west},
            column 2/.style={anchor=west},
            column 3/.style={anchor=west},
            >=latex,->,
            ampersand replacement=\&
          ] (declarematrix) {
\node{\textbf{Block}};
\&
\&
\node{\textbf{Attributes}};
\\
\node{sequence};
\&
\node[smallblock] (task1) {\textsf{B}$_1$};
\node[smallblock,right=\bdist of task1] (task2) {\textsf{B}$_2$};
\draw[outflow,thin] ($(task1.west)-(\outfdist,0)$) -- (task1);
\draw[sequence,->] (task1) -- (task2);
\draw[outflow,thin] (task2) -- ($(task2.east)+(\outfdist,0)$);
\&
\node{
\begin{tabular}{@{}p{\legendw}@{}}
(1) Arbitrary nested blocks $\textsf{B}_1$ and $\textsf{B}_2$
\end{tabular}
};
\\
\node{\begin{tabular}{@{}l@{}}possible\\completion\end{tabular}};
\&
\node[ExclusiveGateway,minimum size=8mm] (eg) {};
\node {~\,\large\textsf{X}};
\draw[outflow,thin] ($(eg.west)-(\outfdist,0)$) -- (eg);
\draw[outflow,thin,->] (eg) -- node[above] {$\varphi_1$} ($(eg.east)+(2*\outfdist,0)$);
\node[EndEvent,draw,minimum size=.6cm,right=\bdist of eg,yshift=-8mm] (err) {e};
\draw[sequence,->,rounded corners=5pt] (eg) |- node[above right] {$\varphi_2$} (err);
\&
\node{
\begin{tabular}{@{}p{\legendw}@{}}
(1) Conditions $\varphi_1$ and $\varphi_2$\\
(2) Type of end event $e$ (error,msg,none)
\end{tabular}
};
\\
\node{\begin{tabular}{@{}l@{}}deferred\\choice\\ / parallel\end{tabular}};
\&
\node[ExclusiveGateway,minimum size=8mm] (eg) {};
\node {~\,\large{$g$}};
\draw[outflow,thin] ($(eg.west)-(\outfdist,0)$) -- (eg);
\node[smallblock,above right=\bdist of eg] (task1) {\textsf{B}$_1$};
\node[smallblock,below right=\bdist of eg] (task2) {\textsf{B}$_2$};
\node[ExclusiveGateway,minimum size=8mm,below right=\bdist of task1] (eg2) {};
\node[below right=\bdist of task1,xshift=-.5mm,yshift=.5mm] {\large{$g$}};
\draw[outflow,thin] (eg2) -- ($(eg2.east)+(\outfdist,0)$);
\draw[sequence,->,rounded corners=5pt] (eg) |- (task1);
\draw[sequence,->,rounded corners=5pt] (eg) |- (task2);
\draw[sequence,->,rounded corners=5pt] (task1) -| (eg2);
\draw[sequence,->,rounded corners=5pt] (task2) -| (eg2);
\&
\node{
\begin{tabular}{@{}p{\legendw}@{}}
(1) Gateway type $g$: $\textsf{X}$ (def.~choice), $\textsf{+}$ (parallel)\\
(2) Arbitrary nested blocks $\textsf{B}_1$ and $\textsf{B}_2$
\end{tabular}
};
\\
\node{\begin{tabular}{@{}l@{}}choice\end{tabular}};
\&
\node[ExclusiveGateway,minimum size=8mm] (eg) {};
\node {~\,\large{$g$}};
\draw[outflow,thin] ($(eg.west)-(\outfdist,0)$) -- (eg);
\node[smallblock,above right=\bdist of eg] (task1) {\textsf{B}$_1$};
\node[smallblock,below right=\bdist of eg] (task2) {\textsf{B}$_2$};
\node[ExclusiveGateway,minimum size=8mm,below right=\bdist of task1] (eg2) {};
\node[below right=\bdist of task1,xshift=-.5mm,yshift=.5mm] {\large{$g$}};
\draw[outflow,thin] (eg2) -- ($(eg2.east)+(\outfdist,0)$);
\draw[sequence,->,rounded corners=5pt] (eg) |- node[below left] {$\varphi_1$} (task1);
\draw[sequence,->,rounded corners=5pt] (eg) |- node[above left] {$\varphi_2$} (task2);
\draw[sequence,->,rounded corners=5pt] (task1) -| (eg2);
\draw[sequence,->,rounded corners=5pt] (task2) -| (eg2);
\&
\node{
\begin{tabular}{@{}p{\legendw}@{}}
(1) Gateway type $g$: $\textsf{X}$/$\textsf{O}$ (excl./incl.~choice)\\
(2) Conditions $\varphi_1$ and $\varphi_2$\\
(3) Arbitrary nested blocks $\textsf{B}_1$ and $\textsf{B}_2$
\end{tabular}
};
\\
\node{loop};
\&
\node[ExclusiveGateway,minimum size=8mm] (eg) {};
\node {~\,\large{\textsf{X}}};
\draw[outflow,thin] ($(eg.west)-(\outfdist,0)$) -- (eg);
\node[smallblock,above right=\bdist of eg] (task1) {\textsf{B}$_1$};
\node[smallblock,below right=\bdist of eg] (task2) {\textsf{B}$_2$};
\node[ExclusiveGateway,minimum size=8mm,below right=\bdist of task1] (eg2) {};
\node[below right=\bdist of task1,xshift=-.5mm,yshift=.5mm] {\large{\textsf{X}}};
\draw[outflow,thin] (eg2) -- node[above] {$\varphi_1$} ($(eg2.east)+(\outfdist,0)$);
\draw[sequence,->,rounded corners=5pt] (eg) |- (task1);
\draw[sequence,->,rounded corners=5pt] (task2) -| (eg);
\draw[sequence,->,rounded corners=5pt] (task1) -| (eg2);
\draw[sequence,->,rounded corners=5pt] (eg2) |- node[above right] {$\varphi_1$} (task2);
\&
\node{
\begin{tabular}{@{}p{\legendw}@{}}
(1) Conditions $\varphi_1$ and $\varphi_2$\\
(2) Arbitrary nested blocks $\textsf{B}_1$ and $\textsf{B}_2$
\end{tabular}
};
\\
\node{\begin{tabular}{@{}l@{}}event-driven\\choice\end{tabular}};
\&
\node[EventBasedGateway,draw,minimum size=8mm] (eg) {};
\node[draw,regular polygon,regular polygon sides=5,minimum width=2mm,scale=0.6,xshift=5.2mm]{};
\draw[outflow,thin] ($(eg.west)-(\outfdist,0)$) -- (eg);
\node[smallblock,above right=\bdist of eg] (e1) {\textsf{E}$_1$};
\node[smallblock,below right=\bdist of eg] (e2) {\textsf{E}$_2$};
\node[smallblock,right=\bdist of e1] (task1) {\textsf{B}$_1$};
\node[smallblock,right=\bdist of e2] (task2) {\textsf{B}$_2$};
\node[ExclusiveGateway,minimum size=8mm,below right=\bdist of task1,xshift=-2mm] (eg2) {};
\node[below right=\bdist of task1,xshift=-2.5mm,yshift=.5mm] {\large{\textsf{X}}};
\draw[outflow,thin] (eg2) -- ($(eg2.east)+(\outfdist,0)$);
\draw[sequence,->,rounded corners=5pt] (eg) |- (e1);
\draw[sequence,->,rounded corners=5pt] (eg) |- (e2);
\draw[sequence,->,rounded corners=5pt] (e1) -- (task1);
\draw[sequence,->,rounded corners=5pt] (e2) -- (task2);
\draw[sequence,->,rounded corners=5pt] (task1) -| (eg2);
\draw[sequence,->,rounded corners=5pt] (task2) -| (eg2);
\&
\node{
\begin{tabular}{@{}p{\legendw}@{}}
(1) Cath event nested blocks $\textsf{E}_1$ and $\textsf{E}_2$\\
(2) Arbitrary nested blocks $\textsf{B}_1$ and $\textsf{B}_2$
\end{tabular}
};
\\
};
\end{tikzpicture}
}
\caption{Flow \dab blocks; for simplicity, we consider only two nested blocks, but multiple nested blocks can be seamlessly handled.}
\label{fig:flow-blocks}
\end{figure}

\begin{figure}[t]
\centering
\resizebox{\textwidth}{!}{
\begin{tikzpicture}[x=1.8cm,y=.9cm, thick]

\matrix[  nodes={node distance=\bldist},
            rectangle,
            nodes in empty cells,
            row sep=.5mm,column sep=1mm,
            very thick,
            column 1/.style={anchor=west},
            column 2/.style={anchor=west},
            column 3/.style={anchor=west},
            >=latex,->,
            ampersand replacement=\&
          ] (declarematrix) {
\node{\textbf{Block}};
\&
\&
\node{\textbf{Attributes}};
\\
\node{\begin{tabular}{@{}l@{}}backward\\exception\end{tabular}};
\&
\node[ExclusiveGateway,minimum size=8mm] (eg) {};
\node {~\,\large{\textsf{X}}};
\draw[outflow,thin] ($(eg.west)-(\outfdist,0)$) -- (eg);
\node[smallblock,right=2*\bdist of eg] (task) {\textsf{A}};
\draw[sequence,->] (eg) -- (task);
\draw[outflow,thin] (task) -- ($(task.east)+(\outfdist,0)$);
\node[IntermediateEvent,minimum size=6mm,right=0mm of task,anchor=center,yshift=-5mm,fill=white] (be) {$e$};
\node[smallblock,below right=\bdist of eg] (b) {\textsf{B}};
\draw[sequence,->,rounded corners=5pt] (be) |- (b);
\draw[sequence,->,rounded corners=5pt] (b) -| (eg);
\&
\node{
\begin{tabular}{@{}p{\legendw}@{}}
(1) Type of boundary event $e$ (error,msg,timer)\\
(2) Subprocess nested block $\textsf{A}$\\
(3) Arbitrary nested block $\textsf{B}$
\end{tabular}
};
\\
\node{\begin{tabular}{@{}l@{}}forward\\exception\end{tabular}};
\&
\node[smallblock] (a) {\textsf{A}};
\draw[outflow,thin] ($(a.west)-(\outfdist,0)$) -- (a);
\node[smallblock,right=\bdist of a] (b1) {\textsf{B}$_1$};
\node[smallblock,below=1mm of b1] (b2) {\textsf{B}$_2$};
\node[ExclusiveGateway,minimum size=8mm,right=\bdist of b1] (eg) {};
\draw[outflow,thin] (eg) -- ($(eg.east)+(\outfdist,0)$);
\node[right=\bdist of b1,xshift=-.3mm] {~\,\large{\textsf{X}}};
\draw[sequence,->] (a) -- (b1);
\draw[sequence,->] (b1) -- (eg);
\node[IntermediateEvent,minimum size=6mm,below=0mm of a,anchor=center,fill=white,yshift=-1mm] (be) {$e$};
\draw[sequence,->,rounded corners=5pt] (be) |- (b2);
\draw[sequence,->,rounded corners=5pt] (b2) -| (eg);
\&
\node{
\begin{tabular}{@{}p{\legendw}@{}}
(1) Type of boundary event $e$ (error,msg,timer)\\
(2) Subprocess nested block $\textsf{A}$\\
(3) Arbitrary nested blocks $\textsf{B}_1$ and $\textsf{B}_2$
\end{tabular}
};
\\
\node{\begin{tabular}{@{}l@{}}forward\\non-interrupting\\exception\end{tabular}};
\&
\node[smallblock] (a) {\textsf{A}};
\draw[outflow,thin] ($(a.west)-(\outfdist,0)$) -- (a);
\node[smallblock,right=\bdist of a] (b1) {\textsf{B}$_1$};
\node[smallblock,below=1mm of b1] (b2) {\textsf{B}$_2$};
\node[ExclusiveGateway,minimum size=8mm,right=\bdist of b1] (eg) {};
\draw[outflow,thin] (eg) -- ($(eg.east)+(\outfdist,0)$);
\node[right=\bdist of b1,xshift=-.3mm] {~\,\large{\textsf{O}}};
\draw[sequence,->] (a) -- (b1);
\draw[sequence,->] (b1) -- (eg);
\node[IntermediateEvent,minimum size=6mm,below=0mm of a,anchor=center,fill=white,yshift=-1mm,densely dashed] (be) {$e$};
\draw[sequence,->,rounded corners=5pt] (be) |- (b2);
\draw[sequence,->,rounded corners=5pt] (b2) -| (eg);
\&
\node{
\begin{tabular}{@{}p{\legendw}@{}}
(1) Type of boundary event $e$ (msg,timer)\\
(2) Subprocess nested block $\textsf{A}$\\
(3) Arbitrary nested blocks $\textsf{B}_1$ and $\textsf{B}_2$
\end{tabular}
};
\\
};
\end{tikzpicture}
}
\caption{\dab exception handling blocks; for simplicity, we show a single boundary event, but multiple boundary events and their corresponding handlers can be seamlessly handled.}
\label{fig:exception-blocks}
\end{figure}
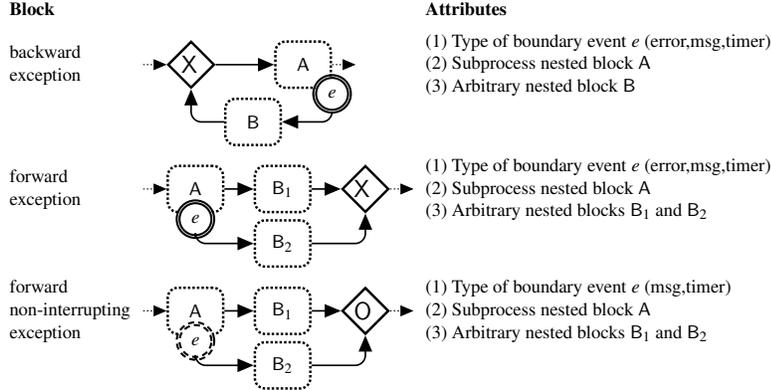

\subsection{Process Schema}
A process schema consists of a block-structured BPMN diagram, enriched with conditions and update effects expressed over a given data schema, according to what described in the previous sections. As for the control flow, we consider a wide range of block-structured patterns compliant with the standard. 
We focus on private BPMN processes, thereby handling incoming messages in a pure nondeterministic way. So we do for timer events, nondeterministically accounting for their expiration without entering into their metric temporal semantics. Focusing on block-structured components helps us in obtaining a direct,  execution semantics, and a consequent modular and clean translation of various BPMN constructs (including boundary events and exception handling). However, it is important to stress that our approach would seamlessly work also for non-structured processes where each case introduces boundedly many tokens. 

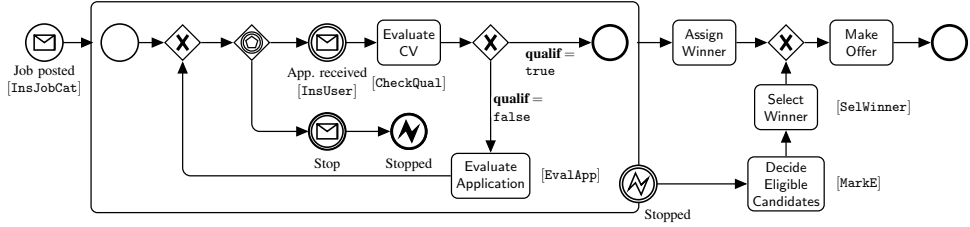
\begin{figure}[t!]
\hspace*{-1.3cm}
\resizebox{1.2\textwidth}{!}{
\begin{tikzpicture}[auto,x=1.4cm,y=.9cm, thick,minimum size=.8cm]

\node[draw,rectangle,rounded corners,minimum width = 12.2cm,minimum height=4.7cm] (spb) at (6.6,-1.6) {} ;

\node[MessageStartEvent] (jobe) at (1.5,0) {};
\node[lbl,below of=jobe,align = center] (jobe_name) {Job posted\\ $\left[\updatespec{InsJobCat}\right]$\\~};
\draw[sequence,->]  (jobe) -- (2.28,0);

\node[StartEvent] (se1) at (2.7,0) {};
\node[lbl,below of=se1,align = center] (se1_name) {};
\node[ExclusiveGateway,draw] (eg1) at (3.7,0) {};
\draw[sequence,->]  (se1) -- (eg1);

\node[EventBasedGateway,draw] (eg_err) at (4.8,0) {};
\node[draw,regular polygon,regular polygon sides=5,minimum width=1mm,scale=0.35] at (4.8,0){};
\draw[sequence,->]  (eg1) -- (eg_err);

\node[MessageIntermediateCatchingEvent] (appe) at (6,0) {};
\node[lbl ,below of=appe,align = center,yshift=1mm] (appe_name) {App.~received\\$\left[\updatespec{InsUser}\right]$};
\draw[sequence,->]  (eg_err) -- 
(appe);

\node[MessageIntermediateCatchingEvent,draw,minimum size=.8cm] (msg-exp) at (6,-2.2) {};
\node[lbl ,below of=msg-exp,align = center,yshift=1mm] (msg-exp_name) {Stop\\~};
\draw[sequence,rounded corners=5pt,->] (eg_err) |- 
(msg-exp);

\node[ErrorEndEvent,draw,minimum size=.8cm] (err) at (7.3,-2.2) {};
\node[lbl ,below of=err,align = center,yshift=1mm] (msg-exp_name) {Stopped\\~};
\draw[sequence,rounded corners=5pt,->] (msg-exp)--(err);


\node[task,align=center] (evalcv) at (7.3,0) {\tname{Evaluate}\\\tname{CV}};
\node[lbl ,below of=evalcv,align = center,yshift=1mm] {$\left[\updatespec{CheckQual}\right]$};
\draw[sequence,->]  (appe) -- (evalcv);

\node[ExclusiveGateway,draw] (eg2) at (8.6,0) {};
\draw[sequence,->]  (evalcv) -- (eg2);

\node[EndEvent] (ee1) at (10.5,0) {};
\node[lbl,below of=ee1,align = center] (ee1_name) {};
\draw[sequence,->]  (eg2) -- node[below,align=left] {$\cvar{qualif}=$\\\true} (ee1);

\node[task,align=center] (evalapp) at (8.6,-3.3) {\tname{Evaluate}\\\tname{Application}};
\node[lbl ,right of=evalapp,anchor=west,align = left] {$\left[\updatespec{EvalApp}\right]$};
\draw[sequence,->,rounded corners=5pt] (eg2) -- node[guard,align=left] {$\cvar{qualif}=$\\\false} (evalapp);
\draw[sequence,->,rounded corners=5pt] (evalapp) -| (eg1);

\node[ErrorIntermediateEvent,fill=white] (eb) at (10.95,-3.5) {};
\node[lbl,below right of=eb,align = center,xshift=-.05cm] (eb_name) {Stopped};

\node[task,align=center] (eligcand) at (13.3,-3.5) {\tname{Decide}\\\tname{Eligible}\\\tname{Candidates}};
\node[lbl ,right of=eligcand,anchor=west,align = left] {$\left[\updatespec{MarkE}\right]$};
\draw[sequence,->,rounded corners=5pt] (eb) -- (eligcand);

%
%

\node[task,align=center] (dwinner) at (13.3,-1.6) {\tname{Select}\\\tname{Winner}};
\node[lbl ,right of=dwinner,anchor=west,align = left] {$\left[\updatespec{SelWinner}\right]$};
\draw[sequence,->]  (eligcand) --  (dwinner);

%
%
\node[task,align=center] (awinner) at (12,0) {\tname{Assign}\\\tname{Winner}};
\draw[sequence,->,rounded corners=5pt] (10.93,0) -- (awinner);

\node[ExclusiveGateway,draw] (eg3) at (13.3,0) {};
\draw[sequence,->]  (awinner) --  (eg3);
\draw[sequence,->]  (dwinner) --  (eg3);

\node[task,align=center] (offer) at (14.5,0) {\tname{Make}\\\tname{Offer}};
\draw[sequence,->]  (eg3) --  (offer);

\node[EndEvent] (ee2) at (15.9,0) {};
\node[lbl,below of=ee2,align = center] (ee2_name) {};
\draw[sequence,->]  (offer) --  (ee2);

\end{tikzpicture}
}
\caption{The job hiring process. Elements in squared brackets attach the update specifications in Examples~\ref{ex:updates-1} and \ref{ex:updates-2} to corresponding tasks/events.}
\label{fig:bpmn-hiring}
\end{figure}

As usual, blocks are recursively decomposed into sub-blocks, the leaves being task or empty blocks. Depending on its type, a block may come with one or more nested blocks, and be associated with other elements, such as conditions, types of the involved events, and the like. We consider a wide range of blocks, covering basic (cf.~Figure~\ref{fig:basic-blocks}), flow (cf.~Figure~\ref{fig:flow-blocks}), and exception handling (cf.~Figure~\ref{fig:exception-blocks}) patterns. Figure~\ref{fig:bpmn-hiring} gives an idea about what is covered by our approach.
 With these blocks at hand, we finally obtain the full definition of a \dab.

\begin{definition}
A \dab $\M$ is a pair $\tup{\D,\P}$ where $\D$ is a data schema, and $\P$ is a root \emph{process block} such that all conditions and update effects attached to $\P$ and its descendant blocks are expressed over $\D$.
\end{definition}

\begin{example}
  The full hiring job process is shown in Figure~\ref{fig:bpmn-hiring}, using the update effects described in Examples~\ref{ex:updates-1} and \ref{ex:updates-2}. Intuitively, the process works as follows. A case is created when a job is posted, and enters into a looping subprocess where it expects candidates to apply. Specifically, the case waits for an incoming application, or for an external message signalling that the hiring has to be stopped (e.g., because too much time has passed from the posting). Whenever an application is received, the CV of the candidate is evaluated, with two possible outcomes. The first outcome indicates that the candidate directly qualifies for the position, hence no further applications should be considered. In this case, the process continues by declaring the candidate as winner, and making an offer to her. The second outcome of the CV evaluation is instead that the candidate does not directly qualify. A more detailed evaluation is then carried out, assigning a score to the application and storing the outcome into the process repository, then waiting for additional applications to come. When the application management subprocess is stopped (which we model through an error so as to test various types of blocks in the experiments reported in Section~\ref{sec:mcmt}), the applications present in the repository are all processed in parallel, declaring which candidates are eligible and which not depending on their scores. Among the eligible ones, a winner is then selected, making an offer to her. 
  We implicitly assume here that at least one applicant is eligible, but we can easily extend the \dab to account also for the case where no application is eligible.
\end{example}

As customary, each block has a lifecycle that indicates the current state of the block, and how the state may evolve depending on the specific semantics of the block, and the evolution of its inner blocks. In Section~\ref{sec:update-logic} we have already characterized the lifecycle of tasks and catch events. For the other blocks, we continue to use the standard states $\sidle$, $\senabled$, $\sactive$ and $\scompleted$. We use the very same rules of execution described in the BPMN standard to regulate the progression of blocks through such states, taking advantage from the fact that, being the process block-structured, only one instance of a block can be enabled/active at a given time for a given case. For example, the lifecycle of a sequence block $\textsf{S}$ with nested blocks $\textsf{B}_1$ and $\textsf{B}_2$ can be described as follows (considering that the transitions of $\textsf{S}$ from $\sidle$ to $\senabled$ and from $\scompleted$ back to $\sidle$ are inductively regulated by its parent block):
\begin{inparaenum}[\it (i)]
\item if $\textsf{S}$ is $\senabled$, then it becomes $\sactive$, simultaneously inducing a transition of $\textsf{B}_1$ from $\sidle$ to $\senabled$;
\item if  $\textsf{B}_1$ is $\scompleted$, then it becomes $\sidle$, simultaneously inducing a transition of  $\textsf{B}_2$ from $\sidle$ to $\senabled$;
\item if  $\textsf{B}_2$ is $\scompleted$, then it becomes $\sidle$, simultaneously inducing $\textsf{S}$ to move from $\sactive$ to $\scompleted$.
\end{inparaenum}
The lifecycle of other block types can be defined analogously.

\subsection{Execution Semantics}
\label{sec:exec}
We intuitively describe the execution semantics of a case over  \dab $\M = \tup{\D,\P}$, using the update/task logic and progression rules of blocks as a basis. Upon execution, each state of $\M$ is characterized by an \emph{$\M$-snapshot}, in turn constituted by a data snapshot of $\D$ (cf.~Section~\ref{sec:data-schema}), and a further assignment mapping each block in $\P$ to its current lifecycle state.

Initially, the data snapshot fixes the immutable content of the catalog $\dcat{\D}$, while the repository instance is empty, the case assignment is initialized to all $\nullv$, and the control assignment assigns to all blocks in $\P$ the $\sidle$ state, with the exception of $\P$ itself, which is $\senabled$. At each moment in time, the $\M$-snapshot is then evolved by nondeterministically evolving the case through one of the executable steps in the process, depending on the current $\M$-snapshot. If the execution step is about the progression of the case inside the process control-flow, then the control assignment is updated. If instead the execution step is about the application of some update effect, the new $\M$ -snapshot is then obtained by following Section~\ref{sec:update-logic}. 

\newcommand{\tsys}[3]{\Upsilon^{#3}_{#1,#2}}
\newcommand{\bralgo}{\mathsf{BackReach}}

\tikzstyle{arrayel} = [
  minimum width=4mm,
  minimum height=4mm,
  rectangle,
  thick,
  draw
]

\tikzstyle{cell} = [
  minimum width=4mm,
  minimum height=2mm,
  rectangle,
  thick,
  draw,
  font=\tiny
]

\tikzstyle{index} = [
  fill=orange!20,
  rectangle,
  rounded corners=5pt,
  draw=orange!80,
  densely dashed,
  very thick,
]

\tikzstyle{elem} = [
  fill=purple!20,
  rectangle,
  rounded corners=5pt,
  draw=purple!80,
  densely dashed,
  very thick,
]

\tikzstyle{point} = [
  center*->,
  rounded corners=5pt,
  thick,
]

\begin{figure}[t]
\subfloat[][Insertion of value \constant{"s"} into an empty string array\label{fig:array-simple}]{
\begin{tikzpicture}[x=1mm,y=0cm, node distance=2mm, thick,font=\footnotesize]
  \node (i0) {0};
  \node[below=0mm of i0,inner sep=0] (i1) {1};
  \node[below=2.5mm of i1,xshift=.2mm,rotate=90,anchor=base] (ietc) {\ldots};

  \begin{pgfonlayer}{background}
    \node[index,fit=(i0) (ietc),label=below:indexes] (index) {};
  \end{pgfonlayer}

  \node[arrayel,right=3mm of i0,deepgreen!90] (e00) {};
  \node[arrayel, below=-\pgflinewidth of e00,deepgreen!90] (e10) {};
  \node[arrayel, below=-\pgflinewidth of e10,deepgreen!90, densely dashed] (e20) {};
  \node[below=0mm of e20,deepgreen!90] (a0) {$a$};

  \node[arrayel,right=2cm of i0,deepblue!90] (e01) {\constant{s}};
  \node[arrayel, below=-\pgflinewidth of e01,deepblue!90] (e11) {};
  \node[arrayel, below=-\pgflinewidth of e11,deepblue!90] (e21) {};
  \node[below=0mm of e21,deepblue!90] (a1) {$a$};

  \draw[|->,thick] (a0) -- node[below,font=\tiny] {insert \constant{"s"} into $a$} (a1);

  \node[below=3mm of a1] {}; 
\end{tikzpicture}
}
\hfill
\subfloat[][Array-based representation of the job hiring repository of Example~\ref{ex:data-schema}, and manipulation of a job application with a fixed catalog. \label{fig:array-repository}]{
\begin{tikzpicture}[x=1mm,y=0cm, node distance=2mm, thick,font=\footnotesize]
  \node (i0) {0};
  \node[below=0mm of i0,inner sep=0] (i1) {1};
  \node[below=2.5mm of i1,xshift=.2mm,rotate=90,anchor=base] (ietc) {\ldots};
   
  \node[right=1.5cm of i0,yshift=20mm,font=\scriptsize] (jc) {\relname{JobCategory}};
  \node[cell,below=0mm of jc,minimum width=12mm] (jc1) {\constant{analyst}}; 
  \node[cell,below=-\pgflinewidth of jc1,minimum width=12mm] (jc2) {\constant{programmer}}; 
  
  \node[cell,right=7.2mm of jc1,minimum width=3mm] (us11) {\constant{u1}}; 
  \node[cell,right=-\pgflinewidth of us11,minimum width=7mm] (us12) {\constant{alice}}; 
  \node[cell,right=-\pgflinewidth of us12,minimum width=3mm] (us13) {\constant{20}}; 
  \node[cell,below=-\pgflinewidth of us11,minimum width=3mm] (us21) {\constant{u2}}; 
  \node[cell,right=-\pgflinewidth of us21,minimum width=7mm] (us22) {\constant{bob}}; 
  \node[cell,right=-\pgflinewidth of us22,minimum width=3mm] (us23) {\constant{23}};  
  \node[cell,below=-\pgflinewidth of us21,minimum width=3mm] (us31) {\constant{u3}}; 
  \node[cell,right=-\pgflinewidth of us31,minimum width=7mm] (us32) {\constant{dana}}; 
  \node[cell,right=-\pgflinewidth of us32,minimum width=3mm] (us33) {\constant{22}};  
  
  \node[above=.5mm of us12,font=\scriptsize] (us) {\relname{User}};

%
  \begin{pgfonlayer}{background}
    \node[index,fit=(i0) (ietc),label=below:indexes] (index) {};
    \node[elem,fit=(jc) (us33),label=above:catalog,inner sep=5pt] (elem) {};
  \end{pgfonlayer}

  \node[arrayel,right=5mm of i0,deepgreen!90] (j00) {};
  \node[arrayel, below=-\pgflinewidth of j00,deepgreen!90] (j10) {};
  \node[arrayel, below=-\pgflinewidth of j10,deepgreen!90,densely dashed] (j20) {};
  \node[below=0mm of j20,deepgreen!90,font=\scriptsize] (j0) {$Jcid$};
  
  \node[arrayel,right=.5mm of j00,deepgreen!90] (u00) {};
  \node[arrayel, below=-\pgflinewidth of u00,deepgreen!90] (u10) {};
  \node[arrayel, below=-\pgflinewidth of u10,deepgreen!90,densely dashed] (u20) {};
  \node[below=0mm of u20,deepgreen!90,font=\scriptsize] (u0) {$Uid$};

  \node[arrayel,right=.5mm of u00,deepgreen!90] (s00) {};
  \node[arrayel, below=-\pgflinewidth of s00,deepgreen!90] (s10) {};
  \node[arrayel, below=-\pgflinewidth of s10,deepgreen!90,densely dashed] (s20) {};
  \node[below=0mm of s20,deepgreen!90,font=\scriptsize] (s0) {$Score$};

  \node[arrayel,right=.5mm of s00,deepgreen!90] (e00) {};
  \node[arrayel, below=-\pgflinewidth of e00,deepgreen!90] (e10) {};
  \node[arrayel, below=-\pgflinewidth of e10,deepgreen!90,densely dashed] (e20) {};
  \node[below=0mm of e20,deepgreen!90,font=\scriptsize] (e0) {$Eli$};

  \node[below=-1mm of u0.south east,deepgreen!90,font=\scriptsize] (a0) {$Application$};

  \node[arrayel,right=3cm of i0,deepblue!90] (j01) {};
  \node[arrayel, below=-\pgflinewidth of j01,deepblue!90] (j11) {};
  \node[arrayel, below=-\pgflinewidth of j11,deepblue!90,densely dashed] (j21) {};
  \node[below=0mm of j21,deepblue!90,font=\scriptsize] (j1) {$Jcid$};
  
  \node[arrayel,right=.5mm of j01,deepblue!90] (u01) {};
  \node[arrayel, below=-\pgflinewidth of u01,deepblue!90] (u11) {};
  \node[arrayel, below=-\pgflinewidth of u11,deepblue!90,densely dashed] (u21) {};
  \node[below=0mm of u21,deepblue!90,font=\scriptsize] (u1) {$Uid$};

  \node[arrayel,right=.5mm of u01,deepblue!90] (s01) {};
  \node[arrayel, below=-\pgflinewidth of s01,deepblue!90] (s11) {};
  \node[arrayel, below=-\pgflinewidth of s11,deepblue!90,densely dashed] (s21) {};
  \node[below=0mm of s21,deepblue!90,font=\scriptsize] (s1) {$Score$};

  \node[arrayel,right=.5mm of s01,deepblue!90] (e01) {};
  \node[arrayel, below=-\pgflinewidth of e01,deepblue!90] (e11) {};
  \node[arrayel, below=-\pgflinewidth of e11,deepblue!90,densely dashed] (e21) {};
  \node[below=0mm of e21,deepblue!90,font=\scriptsize] (e1) {$Eli$};

  \node[below=-1mm of u1.south east,deepblue!90,font=\scriptsize] (a1) {$Application$};

  \node[arrayel,right=5.5cm of i0,violet!90] (j02) {};
  \node[arrayel, below=-\pgflinewidth of j02,violet!90] (j12) {};
  \node[arrayel, below=-\pgflinewidth of j12,violet!90,densely dashed] (j22) {};
  \node[below=0mm of j22,violet!90,font=\scriptsize] (j2) {$Jcid$};
  
  \node[arrayel,right=.5mm of j02,violet!90] (u02) {};
  \node[arrayel, below=-\pgflinewidth of u02,violet!90] (u12) {};
  \node[arrayel, below=-\pgflinewidth of u12,violet!90,densely dashed] (u22) {};
  \node[below=0mm of u22,violet!90,font=\scriptsize] (u2) {$Uid$};

  \node[arrayel,right=.5mm of u02,violet!90] (s02) {\tiny\constant{85}};
  \node[arrayel, below=-\pgflinewidth of s02,violet!90] (s12) {};
  \node[arrayel, below=-\pgflinewidth of s12,violet!90,densely dashed] (s22) {};
  \node[below=0mm of s22,violet!90,font=\scriptsize] (s2) {$Score$};

  \node[arrayel,right=.5mm of s02,violet!90] (e02) {};
  \node[arrayel, below=-\pgflinewidth of e02,violet!90] (e12) {};
  \node[arrayel, below=-\pgflinewidth of e12,violet!90,densely dashed] (e22) {};
  \node[below=0mm of e22,violet!90,font=\scriptsize] (e2) {$Eli$};

  \node[below=-1mm of u2.south east,violet!90,font=\scriptsize] (a2) {$Application$};

  \draw[|->,thick] 
    (a0) 
    -- 
    node[below,font=\tiny] {\constant{u3} applies as \constant{programmer}}
    (a1);
  
    \draw[|->,thick] (a1) -- node[below,font=\tiny] {her application scores $85$} (a2);

  \draw[point,deepblue!90] ($(j01.center)$) -| ($(jc2.south east)-(4mm,0)$);
  \draw[point,deepblue!90] ($(u01.center)$) -- ($(us31.south)-(1mm,0)$);
  
  \draw[point,violet!90] 
    ($(j02.center)$) 
    |- 
    ($(j01.north)+(0,1.5mm)$) 
    -| 
    ($(jc2.south east)-(2mm,0)$);

  \draw[point,violet!90] 
    ($(u02.center)$) 
    |- 
    ($(j02.north)+(0,3mm)$) 
    -| 
    ($(us31.south)+(1mm,0)$);

\end{tikzpicture}
}
\caption{Graphical intuition showing the evolution of different array-based systems. The current state of the array is represented in green, whereas consequent states resulting from updates are shown in blue and violet. Empty cells implicitly hold the $\nullv$ value of their corresponding type.}

\end{figure}
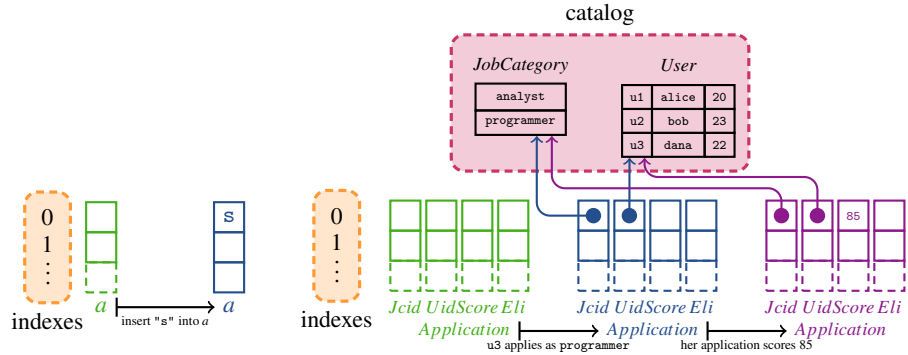


\section{Parameterized Verification of Safety Properties}
\label{sec:verification}
We now focus on parameterized verification of \dab{s} using SMT-based techniques grounded in the theory of arrays. 

\subsection{Array-Based Artifact Systems and Safety Checking}
\label{sec:arrays}
We recall the key notions behind array-based systems, and the array-based artifact systems recently studied in \cite{CGGMR19} to bridge the gap between SMT-based model checking of array-based systems  \cite{ijcar08,lmcs}, and  verification of data- and artifact-centric processes \cite{DeHL18,DeLV16}. 

In general terms, an array-based system logically describes the evolution of array data structures of unbounded size. Figure~\ref{fig:array-simple} intuitively shows a simple array-based system consisting of a single array storing strings. The logical representation of an array relies on a theory with two types of sorts, one accounting for the array indexes, and the other for the elements stored in the array cells. Since the content of an array changes over time, it is referred to using a \emph{function} variable, called \emph{array state variable}. 
The interpretation of such a variable in a state is that of a total function mapping indexes to elements: for each index, it returns the element stored by the array in that index. In the initial green state of Figure~\ref{fig:array-simple}, the array $a$ is interpreted as a total function mapping every index to the undefined string. 

 Starting from an initial configuration, the interpretation changes when moving from one state to another, reflecting the intended manipulation on the array. 
 Hence, the definition of an array-based system with array state variable $a$ always requires
\begin{inparaenum}[\it (i)]
\item a state formula $I(a)$ describing the \emph{initial configuration(s)} of the array $a$;
\item  a formula $\tau(a,a')$ describing the \emph{transitions} that transform the content of the
array from $a$ to $a'$. 
\end{inparaenum}
By suitably using logical operators, $\tau$ can express in a single formula a repertoire of different updates over $a$.

In such a setting, one of the most fundamental, and studied, verification problem is that of checking whether the evolution induced by $\tau$ over $a$ starting from a configuration in $I(a)$ eventually \emph{reaches} one of the \emph{unsafe} configurations described by a state formula $K(a)$. This, in turn, can be tackled by showing that the formula $I(a_0)\wedge \tau(a_0, a_1) \wedge \cdots \wedge \tau(a_{n-1}, a_n)\wedge K(a_n)$ is satisfiable for some $n$. If no such $n$ exists, then no finite run of the system can reach the undesired configurations, and hence the system is safe. Several mature model checkers exist to ascertain (un)safety of these type of systems, such as \textsc{mcmt} \cite{mcmt} 
and \textsc{cubicle}~\cite{cubicle_cav}. Specifically, \mcmt handles this verification problem through a symbolic \emph{backward reachability} procedure. This is a goal-directed procedure that starts from the undesired states captured by $K(a)$, and iteratively computes so-called \emph{preimages}, i.e., logical formulae symbolically describing those states that, through consecutive applications of $\tau$, directly or indirectly reach configurations satisfying $K(a)$. Two checks are then applied, so as to determine whether the procedure has to stop or must continue the iteration. The first one, called \emph{fixpoint check}, tests if the newly computed preimages all coincide with already computed states: if no new state can be produced, the procedure stops by emitting \emph{safe}. Otherwise, a second test, called \emph{fixpoint check}, is applied to determine if  one of such iterated preimages satisfies $I(a)$: if so, the procedure stops by emitting \emph{unsafe} as a verdict; if not, new iterated preimages are computed and the procedure is repeated.  \mcmt generates the proof obligations arising from safety and fixpoint checks, and passes them to a state-of-the-art SMT solver (currently, \textsc{Yices}~\cite{Dutertre} is employed).


In \cite{CGGMR19}, we have extended array-based systems towards an array-based version of the artifact-centric approach, considering in particular the sophisticated model in \cite{verifas}. 
In the resulting formalism, called RAS, a \emph{relational} artifact system accesses a read-only database with keys and foreign keys (cf.~our \dab catalog). In addition, it operates over a set of relations possibly containing unboundedly many updatable entries (cf.~our \dab repository). 
Figure \ref{fig:array-repository} gives an intuitive idea of how this type of system looks like, using the catalog and repository relations from Example~\ref{ex:data-schema}. Contrast this with the simple array system of Figure \ref{fig:array-simple}. On the one hand, the catalog is treated as a rich, background theory, which can be considered as a more sophisticated version of the element sort in basic array systems. On the other hand, each repository relation is treated as a set of arrays, in which each array accounts for one component of the corresponding repository relation. A tuple in the relation is reconstructed by accessing all such arrays with the same index.  
 In \cite{CGGMR19}, we focus on parameterized (un)safety of RAS, verifying whether there exists an instance of the read-only database such that the artifact system can reach an unsafe configuration.  Since the cells of the arrays may point to identifiers in the catalog, in turn related to other catalog relations via foreign keys, the standard backward reachability procedure needs to be suitably revised \cite{CGGMR19}. In fact, when computing preimage formulae over RAS, existentially quantified ``data" variables may be introduced, breaking the format of state formulae. To restore the key property that the preimage of a state is again represented symbolically as a state formula, such additional quantified variables must be eliminated. Suitable quantifier elimination techniques have been studied in \cite{CGGMR19,cade19} and implemented in the latest version~2.8 of \mcmt, which can now natively handle the verification of RAS. In addition, while the unsafety verification is in general undecidable for RAS, several subclasses with decidable unsafety have been singled out. One of such classes corresponds to RAS operating over arrays whose maximum size is bounded a-priori. A RAS of this type is called SAS (for \emph{simple} artifact systems). All in all, the RAS framework provides a natural foundational and practical basis to formally analyze \dab{s}, which we tackle next.

\subsection{Verification Problems for \dab{s}}
First, we need a language to express unsafety properties over a \dab
$\M = \tup{\D,\P}$. Properties are expressed in a fragment of the \emph{guard} language
of Definition~\ref{def:guard} that queries
repo-relations and case variables as well as the cat-relations that tuples from repo-relations or case variables refer to. Properties also query the
control state of $\P$.  This is done by implicitly extending $\D$ with
additional, special case \emph{control variables} that refer to the lifecycle states
of the blocks in $\P$ (where a block named $B$ gets variable
$\cvar{Blifecycle}$).  Given a snapshot, each such variable is assigned to the
lifecycle state of the corresponding block (i.e., $\sidle$, $\senabled$, and
the like). 
We use $\mathit{F}_\P$ to denote the set of all these additional
case \emph{control} variables.

\begin{definition}
\label{def:property}
A \emph{property} over $\M = \tup{\D,\P}$ is a guard $G$ over $\D$ and the control variables of $\P$, such that every non-case variable in $G$ also appears in a relational atom $R(y_1,\dots,y_n)$, where either $R$ is a repo-relation, or $R$ is a cat-relation and $y_1\in \dcvars{\D}$.
\end{definition}

\begin{example}
By naming $\mathit{HP}$ the root process block of Figure~\ref{fig:bpmn-hiring}, the property  $(\cvar{HPlifecycle} = \constant{completed})$ checks whether some case of the process can terminate. 
This property is \emph{unsafe} for our hiring process, since there is at least one way to evolve the process from the start to the end. 
Since \dab processes are block structured, this is enough to ascertain that the hiring process is \emph{sound}. Property $\cvar{EvalApplifecycle}=\constant{completed} \land \relname{Application}(j,u,s,\true)\land s>100$ describes instead the undesired situation where, after the evaluation of an application, there exists an applicant with score greater than $100$. The hiring process is \emph{safe} w.r.t.~this property (cfr. the \textbf{5th} safe property from Section~\ref{sec:mcmt}).
\end{example}

%

We study unsafety of these properties by considering the general case, and also the one where the repository can store only boundedly many tuples, with a fixed bound. In the latter case, we call the \dab \emph{repo-bounded}.    

\inlinetitle{Translating \dab{s} into Array-Based Artifact Systems}
\label{sec:translation}
Given an unsafety verification problem over a \dab $\M = \tup{\D,\P}$, we encode it as a corresponding unsafety verification problem over a RAS that reconstructs the execution semantics of $\M$. We only provide here the main intuitions behind the translation, which is fully addressed in \cite{CGGMR19-techrep-dab-multicase}. In the translation, $\dcat{\D}$ and $\dcvars{\D}$ are mapped into their corresponding abstractions in RAS (namely, the RAS read-only database and artifact variables, respectively). $\drepo{\D}$ is instead encoded using the intuition of Figure \ref{fig:array-repository}: for each $R \in \drepo{\D}$ and each attribute $a \in \relattrs{R}$, a dedicated array is introduced. Array indexes represent (implicit) identifiers of tuples in $R$, in line with our repository model. To retrieve a tuple from $R$, one just needs to access the arrays corresponding to the different attributes of $R$ with the same index. Finally, case variables are represented using (bounded) arrays of size $1$. On top of these data structures, $\P$ is translated into a RAS transition formula that exactly reconstructs the execution semantics of the blocks in $\P$. 

With this transition in place, we define $\bralgo$ as the backward reachability procedure that:
\begin{inparaenum}[(1)] 
\item takes as input 
\begin{inparaenum}[\it (i)] 
\item a \dab $\M$, 
\item a property $\varphi$ to be verified,
\item a boolean indicating whether $\M$ is repo-bounded or not (in the first case, also providing the value of the bound), and 
\item a boolean indicating whether the semantics for insertion is set or multiset;
\end{inparaenum}
\item translates $\M$ into a corresponding RAS $\widehat{\M}$, and $\varphi$ into a corresponding property $\widehat{\varphi}$ over $\widehat{\M}$ (Definition~\ref{def:property} ensures that $\varphi'$ is indeed a RAS state formula);
\item returns the result produced by the \textsc{mcmt} backward reachability procedure (cf.~Section~\ref{sec:arrays}) on $\widehat{\M}$ and $\widehat{\varphi}$. 
\end{inparaenum}


\subsection{Verification Results}
\label{sec:soundness-completeness}

By exploiting the \dab-to-RAS translation and the formal results in
\cite{CGGMR19}, we are now ready to provide our main technical
contributions. First and foremost: \dab{s} can be correctly verified using $\bralgo$.
\begin{theorem}
  \label{thm:case-bounded}
  $\bralgo$ is sound and complete for checking unsafety of \dab{s}
  that use the multiset or set insertion semantics.
\end{theorem}
Soundness tell us that when $\bralgo$ terminates, it produces a correct answer,
while completeness guarantees that whenever a \dab is unsafe with respect to a
property, then $\bralgo$ detects this.  Hence, $\bralgo$ is a semi-decision
procedure for unsafety.

We study additional conditions on the input \dab for which $\bralgo$ is guaranteed to terminate, then becoming a full decision procedure for unsafety.  The first, unavoidable condition is on the constraints used in the catalog: its foreign keys cannot form referential cycles (where a table directly or indirectly refers to itself). This is in line with \cite{verifas,CGGMR19}. To define acyclicity, we associate to a catalog $\cat$ a characteristic graph $G(\cat)$ that captures
the dependencies between relation schema components induced by primary and
foreign keys. Specifically, $G(\cat)$ is a directed graph such that:
\begin{compactitem}[$\bullet$]
\item for every $R \in \cat$ and every attribute $a \in \relattrs{R}$, the pair
  $\tup{R,a}$ is a node of $G(\cat)$ (and nothing else is a node);
\item $\tup{R_1,a_1} \rightarrow \tup{R_2,a_2}$ if and only if one of the two
  following cases apply:
  \begin{inparaenum}[\itshape (i)]
  \item $R_1=R_2$, $a_1 \neq a_2$, and $a_1 = \relid{R_1}$;
  \item $a_2 = \relid{R_2}$ and $a_1$ is a foreign key referring $R_2$.
  \end{inparaenum}
\end{compactitem}

\begin{definition}
  A \dab is \emph{acyclic} if the characteristic graph of its catalog is so.
\end{definition}


\begin{theorem}
  \label{thm:dec-case-repo-bounded}
  $\bralgo$ terminates when verifying properties over
  repo-bounded and acyclic \dab{s} using the multiset or set insertion
  semantics.
\end{theorem}
If the input \dab is not repo-bounded, acyclicity of the catalog is
not enough: termination requires to carefully control the interplay between the
different components of the \dab. While the required conditions are quite
difficult to grasp at the syntactic level, they can be intuitively understood
using the following \emph{locality principle}: 
whenever the progression of the \dab depends on the repository, it does so only
via a single entry in one of its relations. Hence, direct/indirect comparisons
and joins of distinct tuples within the same or different repository relations
cannot be used. To avoid indirect comparisons/joins, queries cannot mix case
variables and repository relations.

Thus, set insertions cannot be supported, since by definition they require to
compare tuples in the same relation. The next definition is instrumental to
enforce locality.

\begin{definition}
  \label{def:separated-guard}
  A \emph{guard} $G \triangleq q(\vec{x}) \leftarrow \bigvee_{i=1}^n Q_i$ over
  data component $\D$ is \emph{separated} if
  $\getdatavars{Q_i}\cap \getdatavars{Q_j}=\emptyset$ for every $i\neq j$, and each $Q_i$ is of the
  form $\chi \land R(\vec{y}) \land \xi$ (with $\chi$, $R(\vec{y})$, and $\xi$
 optional), where:
  \begin{inparaenum}[\itshape (i)]
  \item $\chi$ is a conjunctive query with filters  only over $\dcat{\D}$, and that
    can employ case variables;
  \item $R \in \drepo{\D}$ is a repo-relation schema;
  \item $\vec{y}$ is a tuple of variables and/or constant objects in
    $\datadom$, such that $\vec{y} \cap \dcvars{\D} = \emptyset$, and
    $\getdatavars{\chi}\cap \vec{y}=\emptyset$;
  \item $\xi$ is a conjunctive query with filters over $\dcat{\D}$ only, that
    possibly mentions variables in $\vec{y}$ but does \emph{not} include any
    case variable, and
    such that $\getdatavars{\chi}\cap\getdatavars{\xi}=\emptyset$.
  \end{inparaenum}
  A \emph{property} is \emph{separated} if it is so as a guard.
\end{definition}

Intuitively, a separated guard consists of two isolated parts: one part $\chi$
inspecting the content of case variables and their relationship with the
catalog, and another part $R(\vec{y}) \land \xi$ retrieving a single tuple
$\vec{y}$ in some repository relation $R$, possibly filtering it through
inspection of the catalog via $\xi$.

\begin{example}
  Consider the refinement
  $\getpre{\updatespec{EvalApp}} \triangleq
  \relname{GetScore}(s:\sort{NumScore}) \leftarrow \xi \land \chi$ of the guard
  $\getpre{\updatespec{EvalApp}}$ from Example~\ref{ex:updates-1}, where
  $\chi:=\relname{User}(\cvar{uid},name,age)$ checks if the
  variables $\tup{\cvar{uid},name,age}$ form a tuple in
  $\relname{User}$, and $\xi:=1 \leq s \land s \leq 100$. This guard is
  separated since $\chi$ and $\xi$ match the requirements of the previous
  definition.
 \end{example}
%

\begin{theorem}
  \label{thm:dec-case-bounded}
  Let $\M$ be an acyclic \dab that uses the multiset insertion semantics, and
  is such that for each update specification $\updatespec{u}$ of $\M$, the
  following holds:
  \begin{compactenum}
  \item If $\getpost{\updatespec{u}}$ is an \emph{insert\&set} rule (with
    explicit $\INSERT$ part), $\getpre{\updatespec{u}}$ is \emph{repo-free};
  \item If $\getpost{\updatespec{u}}$ is a \emph{set} rule (with no $\INSERT$
    part), then either
  \begin{inparaenum}[\itshape (i)]
  \item $\getpre{\updatespec{u}}$ is \emph{repo-free}, or
  \item $\getpre{\updatespec{u}}$ is \emph{separated} and \emph{all} case
    variables appear in the $\SET$ part of $\getpost{\updatespec{u}}$;
  \end{inparaenum}
\item If $\getpost{\updatespec{u}}$ is a \emph{delete\&set} rule, then
  $\getpre{\updatespec{u}}$ is \emph{separated} and \emph{all} case variables
  appear in the $\SET$ part of $\getpost{\updatespec{u}}$;
\item If $\getpost{\updatespec{u}}$ is a \emph{conditional update} rule, then
  $\getpre{\updatespec{u}}$ is \emph{repo-free} and \emph{boolean} (i.e., it returns either \false or the empty tuple), so that
  $\getpost{\updatespec{u}}$ only makes use of the new variables introduced in
  its $\UPDATE$ part (as well as constant objects in $\datadom$).
\end{compactenum}
Then, $\bralgo$ terminates when verifying separated properties over $\M$.
\end{theorem}
Notably, the conditions of Theorem~\ref{thm:dec-case-bounded} represent a
concrete, BPMN-like counterpart of the abstract conditions used in
\cite{verifas} and \cite{CGGMR19} towards decidability.

Specifically, Theorem~\ref{thm:dec-case-bounded} uses two conditions:
\begin{inparaenum}[\itshape (i)]
\item repo-freedom, or
\item the combination of separation with the manipulation of \emph{all} case
  variables at once.
\end{inparaenum}
We now intuitively explain how these conditions substantiate the aforementioned
locality principle.  Overall, the main difficulty is that case variables may be
loaded with data objects extracted from the repository. Hence, the usage of a
case variable may mask an underlying reference to a tuple component stored in
some repo-relation.  Given this, locality demands that no two case variables
can simultaneously hold data objects coming from different tuples in the
repository. At the beginning, this is trivially true, since all case variables
are undefined. A safe snapshot guaranteeing this condition continues to stay so
after an insertion of the form mentioned in point 1 of
Theorem~\ref{thm:dec-case-bounded}: a repo-free precondition ensures that the
repository is not queried at all, and hence trivially preserves
locality. Locality may be easily destroyed by arbitrary set or delete\&set
rules whose precondition accesses the repository. Three aspects have to be
considered to avoid this. First, we have to guarantee that the precondition
does not mix case variables and repo-relations:
Theorem~\ref{thm:dec-case-bounded} does so thanks to separation. Second, we
have to avoid that when the precondition retrieves objects from the repository,
it extracts them from different tuples therein: this is again guaranteed by
separation, since only one tuple is extracted.  A third, subtle situation that
would destroy locality is the one in which the objects retrieved from (the same
tuple in) the repository are only used to assign \emph{a proper subset} of the
case variables: the other case variables could in fact still hold objects previously
retrieved from a \emph{different} tuple in the
repository. Theorem~\ref{thm:dec-case-bounded} guarantees that this never
happens by imposing that, upon a set or delete\&set operation, \emph{all} case
variables are involved in the assignment. Those case variables that get objects
extracted from the repository are then guaranteed to all implicitly point to
the same, single repository tuple retrieved by the separated precondition.

\begin{example}  
 By considering the data and process schema of the hiring process \dab, one can directly show that it obeys to all conditions in Theorem~\ref{thm:dec-case-bounded}, in turn guaranteeing termination of $\bralgo$. For example, rule $\updatespec{EvalApp}$ in Example~\ref{ex:updates-1} matches point 1 since $\getpre{\updatespec{EvalApp}}$ is repo-free. $\updatespec{SelWinner}$ from the same example matches instead point 3, since $\getpre{\updatespec{SelWinner}}$ is trivially separated and \emph{all} case variables appear in the $\SET$ part of $\getpost{\updatespec{SelWinner}}$. 
 \end{example}

\begin{wrapfigure}[12]{r}{1.9cm}
\vspace*{-7mm}
\begin{tabular}{|cc|r|}
\hline  
  \multicolumn{2}{|c|}{\emph{prop.}}  
  &
  \emph{time(s)}\\
\hline
  \parbox[t]{2mm}{
    \multirow{5}{*}{\rotatebox[origin=c]{90}{\textbf{safe}}}
    } 
  & 
  \textbf{1} 
  &
  0.20
  \\
  ~& 
  \textbf{2} 
  &
  5.85
  \\
  & 
  \textbf{3} 
  &
  3.56
  \\
  & 
  \textbf{4} 
  &
  0.03
  \\
  & 
  \textbf{5} 
  &
  0.27
  \\
 \hline
  \parbox[t]{2mm}{
    \multirow{5}{*}{\rotatebox[origin=c]{90}{\textbf{unsafe}}}
    } 
  & 
  \textbf{1} 
  &
  0.18
  \\
  & 
  \textbf{2} 
  &
  1.17
  \\
  & 
  \textbf{3} 
  &
  4.45
  \\
  & 
  \textbf{4} 
  &
  1.43
  \\
  & 
  \textbf{5} 
  &
  1.14
  \\
 \hline 
 \end{tabular}
\end{wrapfigure}
\inlinetitle{First Experiments with MCMT}
\label{sec:mcmt}
We have encoded the job hiring \dab described in the paper into \mcmt, systematically following the translation rules recalled in Section~\ref{sec:translation}, and fully spelled out in \cite{CGGMR19-techrep-dab-multicase} when proving the main theorems of Section~\ref{sec:soundness-completeness}.
%
%
%
%
Running MCMT Version~2.8 (\url{http://users.mat.unimi.it/users/ghilardi/mcmt/}), we have checked the encoding of the job hiring \dab for process termination (which took $0.43$sec), and against five safe and five unsafe properties. For example, the \textbf{1}st \textbf{unsafe} property describes the desired situation in which, after having evaluated an application (i.e., \tname{EvalApp} is completed), there exists at least an applicant with a score greater than $0$. Formally:
$\cvar{EvalApplifecycle}=\constant{completed}  \land Application(j,u,score,e)\land score>0$.
The \textbf{4}th \textbf{safe} property represents instead the situation  in which a winner has been selected after the deadline (i.e., \tname{SelWin} is completed), but the case variable $\cvar{result}$ witnesses that the winner is not an eligible candidate. Formally: $\cvar{SelWinlifecycle}=\constant{completed} \land \cvar{result}=\false$.
 \mcmt returns \safe, witnessing that this configuration is not reachable from the initial states. Additional properties (taken from the table on the right) are described in~\cite{CGGMR19-techrep-dab-multicase}.
 
The table on the right summarizes the obtained, encouraging results, reporting the \mcmt running time in seconds.
The \mcmt specifications containing all the properties to check (together with their intuitive interpretation) are available in~\cite{CGGMR19-techrep-dab-multicase}, and all tests are directly reproducible. Experiments were performed on a machine with Ubuntu~16.04, 2.6\,GHz Intel
Core~i7 and 16\,GB RAM.


\section{Conclusion and Discussion}
We have introduced a data-aware extension of BPMN, called \dab, balancing between expressiveness and verifiability. We have shown that parameterized safety problems over \dab{s} can be correctly tackled by off-the-shelf array-based SMT techniques, and in particular by the backward reachability procedure implemented in the \mcmt model checker. Differently from conventional  process-centric verification, the verification language proposed in this paper supports properties that address both process and data aspects. 

We have then identified classes of \dab{s} suitably controlling the data components and the way the process manipulates it, guaranteeing termination of backward reachability. We have finally shown that a realistic example of \dab can be actually verified by \mcmt with a very promising performance.

There are plenty of avenues for future work. We enumerate the most important ones, considering methodological, foundational, and experimental aspects.

From the methodological point of view, the conditions we have introduced to guarantee termination can be seen as modeling principles for data-aware process designers who aim at making their processes verifiable. The applicability of such principles to real-life processes is an open question, calling for genuine, further research on empirical validation on real-world scenarios, as well as on the definition of guidelines helping modeling and refactoring of arbitrary \dab{s} into fully verifiable ones. Frameworks for the empirical validation of data-aware process models have been recently brought forward \cite{RVPV17}, and can be in fact extended also considering the verifiability factor.  

From the foundational perspective we are interested in equipping \dab{s} with datatypes and corresponding rigid predicates, including arithmetic operators, as done in \cite{DeLV16} for artifact systems. This is promising especially considering that there are plenty of state-of-the-art SMT techniques to handle arithmetics. At the same time, we want to attack the main limitation of our approach, namely that guards and conditions are actually existential formulae, and the only (restricted) form of universal quantification available in the update language is that of conditional updates. Universal guards in transition formulae could be very useful in specifications: for example, they would allow us to specify a branch in a job hiring process that is followed only if \emph{no} applicant satisfies a certain condition. 
The question has been debated since longtime in the literature and the most effective solution so far is the introduction of suitable ``monotonic abstractions" (see \cite{AlGS14} for a survey). Notably, this is already implemented in \mcmt.
 Monotonic abstractions could introduce spurious unsafe traces, and in fact \mcmt warns the user about this (in practice, not so frequent) possibility. 
  An orthogonal, challenging question is how, and to what extent, some of the most recent techniques developed for temporal model checking of artifact-centric systems \cite{DeLV16} can be incorporated in our approach, allowing us to prove more sophisticated properties beyond safety.



From the experimental point of view, while a systematic evaluation is out of scope of this paper, the initial experiments carried out in this paper and \cite{CGGMR19} indicate that the approach is promising. We intend to fully automate the translation from \dab{s} to array-based systems, and to set up a benchmark to evaluate the performance of verifiers for data-aware processes, starting from the examples collected in \cite{verifas}: they are inspired by reference BPMN processes, and consequently should be easily encoded as \dab{s}. 

\bibliographystyle{abbrv}
\bibliography{main-bib}

\end{document}